\documentclass[11pt, oneside]{article}
\pdfoutput=1
\usepackage{geometry}                		
\usepackage{graphicx}	
\usepackage{float}			

\usepackage{amssymb}
\usepackage{amsmath}
\usepackage{bbm}
\usepackage{authblk}
\usepackage{multirow}
\usepackage[T1]{fontenc}
\usepackage[utf8]{inputenc}
\usepackage[font=small,labelfont=bf]{caption}
\usepackage{booktabs}
\usepackage[title]{appendix}

\usepackage{myharv}

\begin{document}
\bibliographystyle{kluwer}
\bibdata{SOFR_BIB}
\title{Short Rate Dynamics: A Fed Funds and SOFR perspective$^{\dag}$}
\date{\today}
\author[1]{Karol Gellert}
\author[2,3,4]{Erik Schl\"ogl}
\affil[1]{University of Technology Sydney, Australia}
\affil[2]{University of Technology Sydney, Australia --- Quantitative Finance Research Centre. \texttt{Erik.Schlogl@uts.edu.au}}
\affil[3]{The African Institute for Financial Markets and Risk Management (AIFMRM), University of Cape Town, South Africa}
\affil[4]{Faculty of Science, Department of Statistics, University of Johannesburg, South Africa}
\renewcommand\Authands{ and }
\maketitle
\begin{abstract}
The Secured Overnight Funding Rate (SOFR) is becoming the main Risk–Free Rate benchmark in US dollars, thus interest rate term structure models need to be updated to reflect the key features exhibited by the dynamics of SOFR and the forward rates implied by SOFR futures. Historically, interest rate term structure modelling has been based on rates of substantially longer time to maturity than overnight, but with SOFR the overnight rate now is the primary market observable. This means that the empirical idiosyncrasies of the overnight rate cannot be ignored when constructing interest rate models in a SOFR–based world.

As a rate reflecting transactions in the Treasury overnight repurchase market, the dynamics of SOFR are closely linked to the dynamics Effective Federal Funds Rate (EFFR), which is the interest rate most directly impacted by US monetary policy target rate decisions. Therefore, these rates feature jumps at known times (Federal Open Market Committee meeting dates), and market expectations of these jumps are reflected in prices for futures written on these rates. On the other hand, forward rates implied by Fed Funds and SOFR futures continue to evolve diffusively. The model presented in this paper reflects the key empirical features of SOFR dynamics and is calibrated to futures prices. In particular, the model reconciles diffusive forward rate dynamics with piecewise constant paths of the target short rate.
\end{abstract}
\vfill
\hrulefill

{\scriptsize $^{\dag}$ The authors thank Leif Andersen for helpful comments on a previous draft of this paper. The usual disclaimers apply.}
\newpage
\section{Introduction}
As the Secured Overnight Funding Rate (SOFR) is currently in the process of becoming the key Risk--Free Rate (RFR) benchmark in US dollars, interest rate term structure models need to be updated to reflect this. Historically, interest rate term structure modelling has been based on rates of substantially longer time to maturity than overnight, either directly as in the LIBOR Market Model,\footnote{See \citeasnoun{TJoF:Milt&Sand&Sond:97}, \citeasnoun{OZ:BGM:97} and \citeasnoun{OZ:MR:97}.} or indirectly, in the sense that even models based on the continuously compounded short rate (i.e., with instantaneous maturity)\footnote{Of these, \citeasnoun{Hul&Whi:90a} is the most prominent example.} are typically calibrated to term rates of longer maturities, with any regard to a market overnight rate at best an afterthought. However, with SOFR this situation is reversed: The overnight rate now is the primary market observable, and term rates (i.e., interest rates for longer maturities) will be less readily available and therefore must be inferred (for example from derivatives prices).

Thus the empirical idiosyncrasies of the overnight rate cannot be ignored when constructing interest rate term structure models in a SOFR--based world, and more than longer term rates, these idiosyncrasies are driven by monetary policy. In this paper we closely examine the dynamics of both SOFR and the closely related and more established Effective Fed Funds Rate (EFFR). We find (already by simple inspection) that models, in which the short rate evolves as a diffusion, can no longer be justified by empirical data. Instead, the primary driver of the short rate is the piecewise flat behaviour of the Federal Open Market Committee (FOMC) policy target rate. Concurrently, we observe that the forward rates associated with the policy target rate evolve in a more diffusive manner. A model which reconciles these two features is the main contribution of this paper.

Modelling the target rate may seem not quite reflective of reality, since the FOMC sets a target range rather than a specific rate. However, we present documentary and empirical evidence that the target rate lives on via the Interest on Excess Reserves (IOER). The IOER acting as the target rate is a deliberate strategy by the Federal Reserve, which has proven much more effective at keeping the EFFR near the policy target.

Another prominent empirical feature of SOFR dynamics, and to a lesser degree EFFR dynamics, is the occurrence of large spikes. The spikes tend to occur at predictable times on the last day of the month and particularly end of quarter and end of year dates. Not all spikes occur on the last day of the month, such as the extreme spike in September 2019. An explanation provided by the Federal Reserve\footnote{see Feds Notes link: https://www.federalreserve.gov/econres/notes/feds-notes/what-happened-in-money-markets-in-september-2019-20200227.htm} for the September 2019 spike is that it occurred on a day on which large corporate tax receipts and Treasury bond expiries caused a sharp imbalance in demand and supply in the repo market. Both the reasons given occur on dates easily obtainable in advance, therefore arguably this spike also could be classified as occurring on a predictable date. Using a similar approach to modelling the target rates, we construct a model for spikes occurring on known dates, where the forward rates associated with those spikes evolve in a diffusive manner.

The literature refers to jumps with deterministic jump times as \emph{stochastic discontinuities}, see for example \citeasnoun{Kim2014}, \citeasnoun{KellerRessel2018}, \citeasnoun{Fontana2019}. The nomenclature reflects the context that the discontinuities are treated as extensions to an existing continuous stochastic model. Our approach is distinctly different in that the discontinuity is the basis of our model for the short rate, while simultaneously the forward rates for maturities beyond the next scheduled jump evolve as a continuous stochastic process.

Specific to SOFR, \citeasnoun{Heitfield2019} model forward rates using a step function, assuming that rates remain constant for all dates between FOMC meetings. This is a static approach for the purposes of calibrating a piecewise flat term structure, similar to our assumption in the calibration section. Most recently \citeasnoun{Andersen2020} provide a SOFR--inspired general spike model to enable the extension of derivative pricing models to spikes. While their paper includes spikes at known times as a special case, our focus is exclusively on short rate discontinuities at known times.

Several papers focus on adapting existing models to SOFR without considering discontinuities. These include \citeasnoun{Mercurio2018} who uses a deterministic SOFR--OIS spread with a short rate model for the OIS. \citeasnoun{Lyashenko2019} propose an extension to the LIBOR Market Model to accommodate the in--arrears setting nature of term rates related to SOFR and overnight benchmark rates in general. \citeasnoun{Skov2020} show that a three--factor Gaussian arbitrage--free Nelson/Siegel model is well suited for the SOFR futures market, but they do not include the time series of SOFR itself in their estimation.

The rest of the paper is organised as follows. Section 2 closely examines the empirical behaviour of SOFR and EFFR, which motivates the model proposed in this paper. Section 3 presents the model for discontinuous short rates with continuous forward rates including both step and spike discontinuities. The model is presented within the \citeasnoun{Hea&Jar&Mor:92} (HJM) framework, and also includes a Gaussian diffusion to account for residual noise. Results from calibration of the model to futures market data are presented in Section 4.

\section{Empirical Motivation}\label{sec:empirical}
\subsection{Monetary Policy and Short Rate Models}
Over the course of the last five years significant changes to the implementation of monetary policy have had a dramatic impact on the EFFR, resulting in a substantial divergence between its empirical behaviour and the dynamic assumptions of short rate models. The changes trace back to the 2008 financial crisis, prior to which monetary policy was administered primarily by direct intervention in the Fed Funds market to maintain the EFFR close to the target rate set by the FOMC. The approach relied on open market operations by the Federal Reserve trading desk resulting in the EFFR gravitating around the target rate with varying degrees of volatility.\footnote{See \citeasnoun{Hilton2005} for an analysis of factors impacting EFFR volatility related to open market operations}

\begin{figure}[t]
\centerline{\includegraphics[scale=0.65]{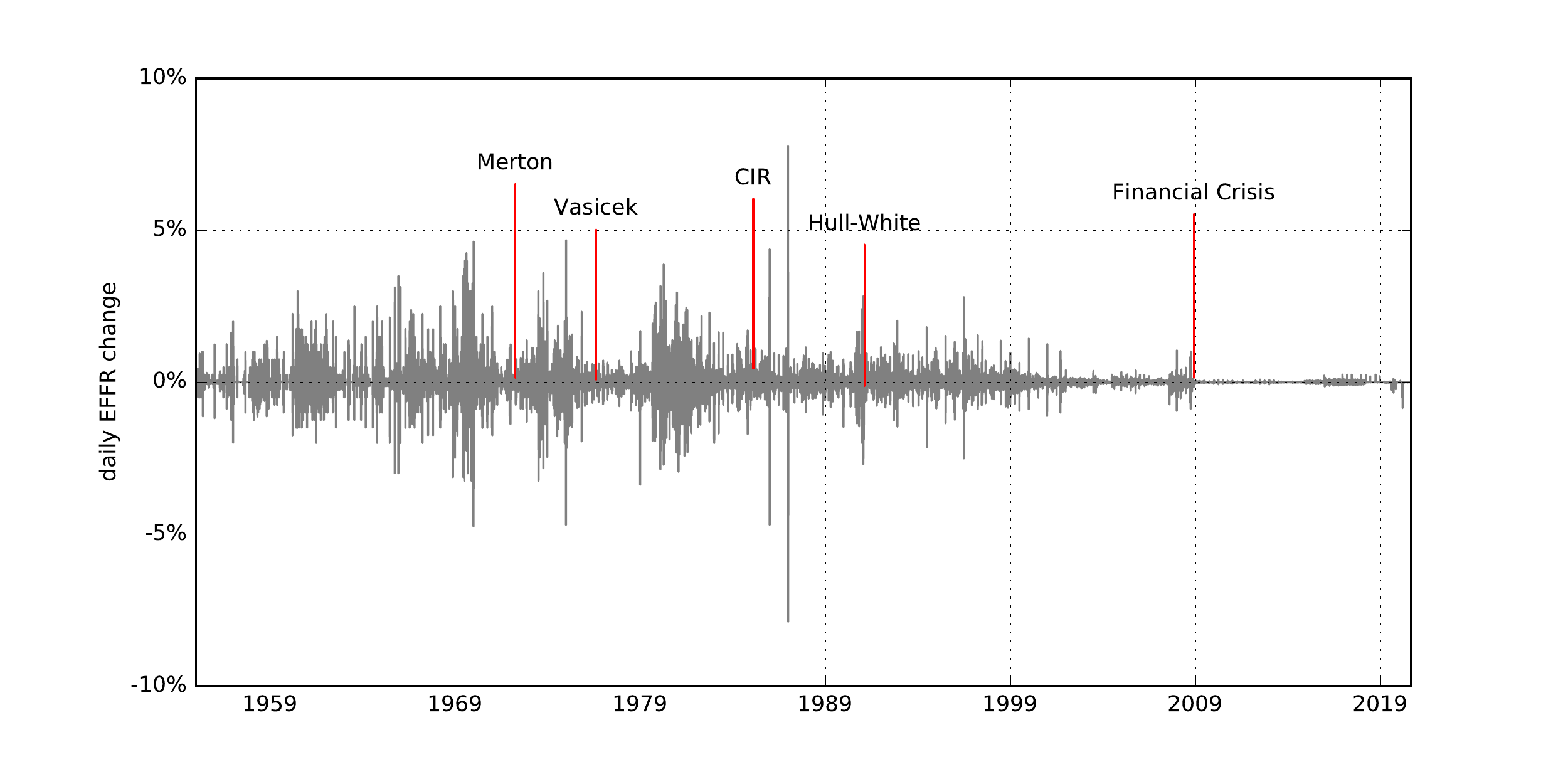}}
\caption{Empirical daily EFFR changes and the history of short rate models}\label{model_timeline}
\end{figure}

The first stochastic model of the short rate is attributed to \citeasnoun{Merton1973} who employed a single--dimension Brownian motion as the driver. At least on cursory visual inspection, the empirical data at the time, see Fig. \ref{model_timeline}, did not contradict the mathematically tractable Gaussian assumption of the model. The next major development came from \citeasnoun{Vasicek1977}, adding mean reversion, a strong empirical feature of rate dynamics. Modelling mean reversion also aligned with the notion of open market operations by the Federal Reserve trading desk managing the rate around the monetary policy target. \citeasnoun{Cox&Ing&Ros:85b} (CIR) modified the dynamics of the continuously compounded short rate by scaling the volatility by the square root of the short rate, ensuring non-negativity of interest rates. The next milestone in short rate modelling was an extension of the Vasicek model with time dependent drift by \citeasnoun{Hul&Whi:90a}, allowing the model to be fitted to an initial term structure of interest rates observed in the market --- this was critical for use of the model to price interest rate derivatives. \citeasnoun{Hea&Jar&Mor:92} developed the general framework into which all diffusion--based arbitrage--free interest rate term structure models must fit.

Open market operations are carried out by the Federal Reserve trading desk whose trading goal is to maintain the EFFR near the target rate. This involves monitoring the market and counteracting trades which move the EFFR away from target, in essence micro--managing market liquidity. The 2008 financial crisis included a crisis in liquidity and the ability of the Federal Reserve's trading desk to maintain the EFFR near the target rate significantly deteriorated. The trading desk did not have the means to counteract the dramatic drain in supply of desperately demanded capital.

\begin{figure}[t]
\centerline{\includegraphics[scale=0.65]{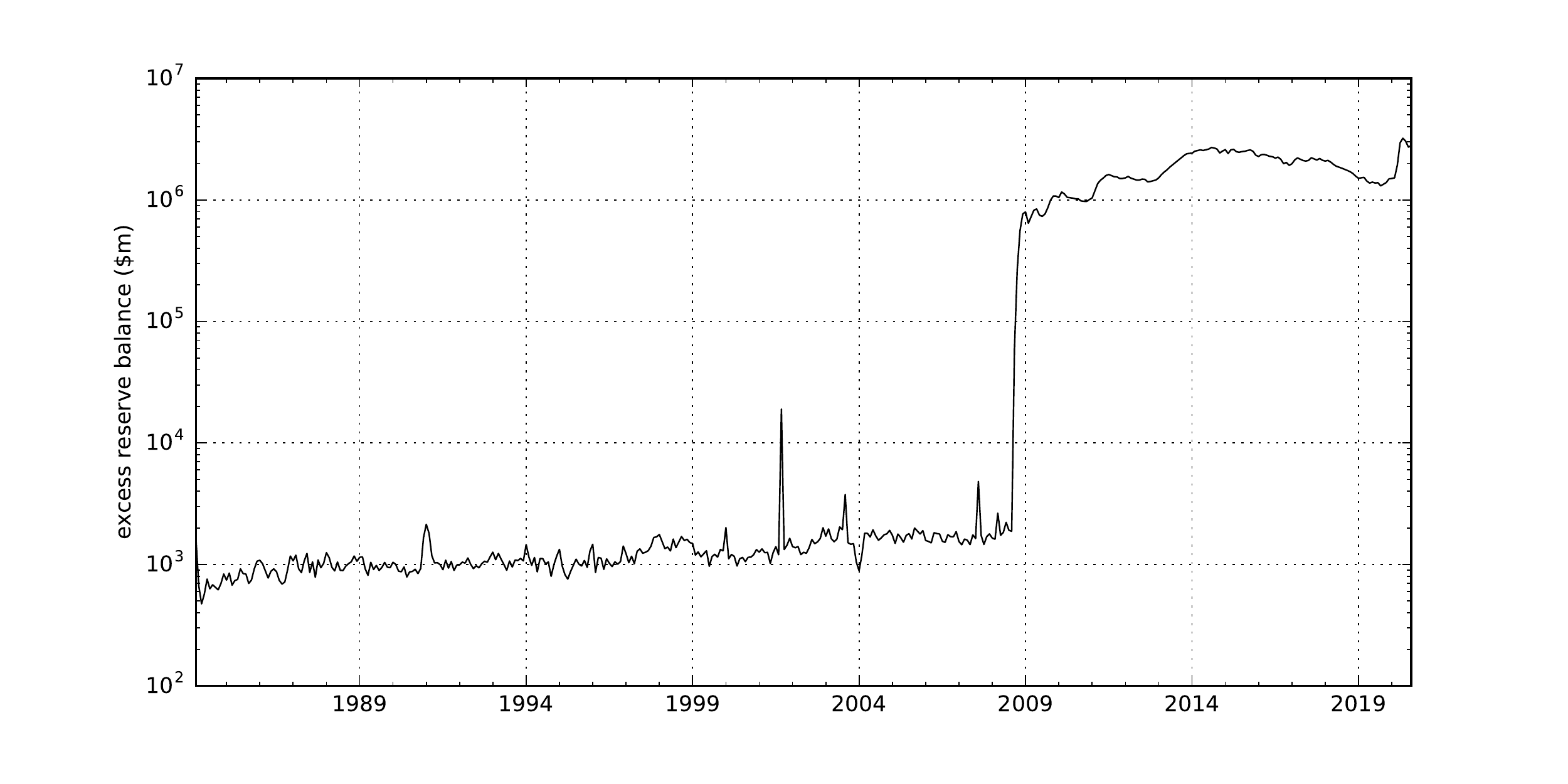}}
\caption{Excess reserves balance history}\label{excess_reserves}
\end{figure}

This was acknowledged by the Federal Reserve\footnote{See \citeasnoun{fomc} December 2008, page 9.} as one of the factors considered when switching to a target range, initially set between 0 and 25 basis points. The Federal Reserve's strategy in response to the financial crisis centred around two key policies: near zero interest rates and quantitative easing. The phases of quantitative easing became known as QE 1/2/3 and involved selling Treasury bonds and purchases of various credit risky assets\footnote{Such as Agency Debt, Mortgage Backed Securities and Term Auction Facilities, see \citeasnoun{Binder2010}.} in a bid to boost liquidity and credit conditions. The Federal Reserve's injection of liquidity resulted in an environment of elevated excess reserves. By historical standards, the rise in excess reserves was extreme and without precedent. As can seen in Fig. \ref{excess_reserves}, it increased from under \$2 billion in September 2008 to \$1 trillion by November 2009, before reaching a high of over \$2.5 trillion in October 2015.

In October 2008, the Federal Reserve began paying IOER\footnote{See \citeasnoun{fomc} October 2008, page 7.} to help control the EFFR in response to increasing excess reserves. It was thought at the time that the IOER should act as a lower bound for the EFFR, since no institutions should want to lend below this rate. As such, effective from October 9 the IOER was set to 75 basis points, with the EFFR target rate at 150 basis points. In the following days the EFFR was setting well below the target rate, including some days below the IOER. On the October 23, to lift rates closer to target, IOER was increased to 110 basis points, in response EFFR rates increased but were still setting below the IOER. Other adjustments were made in November under the assumption of IOER acting as a lower bound, however with EFFR persisting to settle well below the IOER it became clear the assumption was incorrect.

In the FOMC immediately following the introduction of the IOER, it was noted that institutions not eligible to receive it were willing to sell (lend) funds at rates below the IOER.\footnote{See \citeasnoun{fomc} October 2008, page 2.} However, it was not until December 2008, where together with the introduction of the target range, the IOER was set at the target range upper limit of 25 basis points in recognition that due to unique circumstances the IOER was acting as an upper bound for the EFFR. The large surpluses in excess reserves eliminated demand for reserve loans. Instead the Fed Funds rate was driven by Government Sponsored Institutions who do not earn interest on reserve balances, lending their excess reserves at below the IOER to institutions who would then earn the difference between the Fed Funds rate and the IOER. In effect, by paying the IOER in a market flooded with liquidity, the Federal Reserve became the borrower, rather than the lender of last resort.

Plans for reversal of the post financial crisis expansionary policy were formally laid out at the FOMC September 2014 meeting as the Policy Normalization Principles and Plans.\footnote{See \citeasnoun{fomc} September 2014, page 3.} The aim of the normalisation strategy was to bring the EFFR back to normal levels and reduce the securities held by the Federal Reserve, thereby unwinding the excess reserves held by banks. Prior to the financial crisis, controlling the supply of reserves via open market operations was a key tool in controlling the Fed Funds rate. However, the Federal Reserve has adopted the view that with banks using reserves for liquidity more than prior to the crisis, it might be hard to predict demand for reserves and therefore open market operations would not be effective at precisely controlling the EFFR.\footnote{See \citeasnoun{fomc} November 2018, page 3.} Instead, the new normal will constitute the Federal Reserve keeping excess reserves just large enough to remain on the flat part of the demand curve, a prerequisite condition for the use of the IOER to control the EFFR.

Thus the conditions in the Fed Funds market are dramatically different to when short rate models were first conceived. The flood of liquidity in excess reserves, by construction aimed at removing any supply--demand gradient, has removed most of the volatility from the short rate of interest, with changes in the short rate being mainly driven by changes in the IOER, leading to jumps at known times (the FOMC meeting dates). Forward rates implied by traded market instruments, however, continue to exhibit volatility, as the evolution of market expectations of FOMC actions is priced into forward--looking instruments such as Fed Fund futures.

\subsection{Effective Federal Funds Rate}
\begin{figure}[t]
\centerline{\includegraphics[scale=0.65]{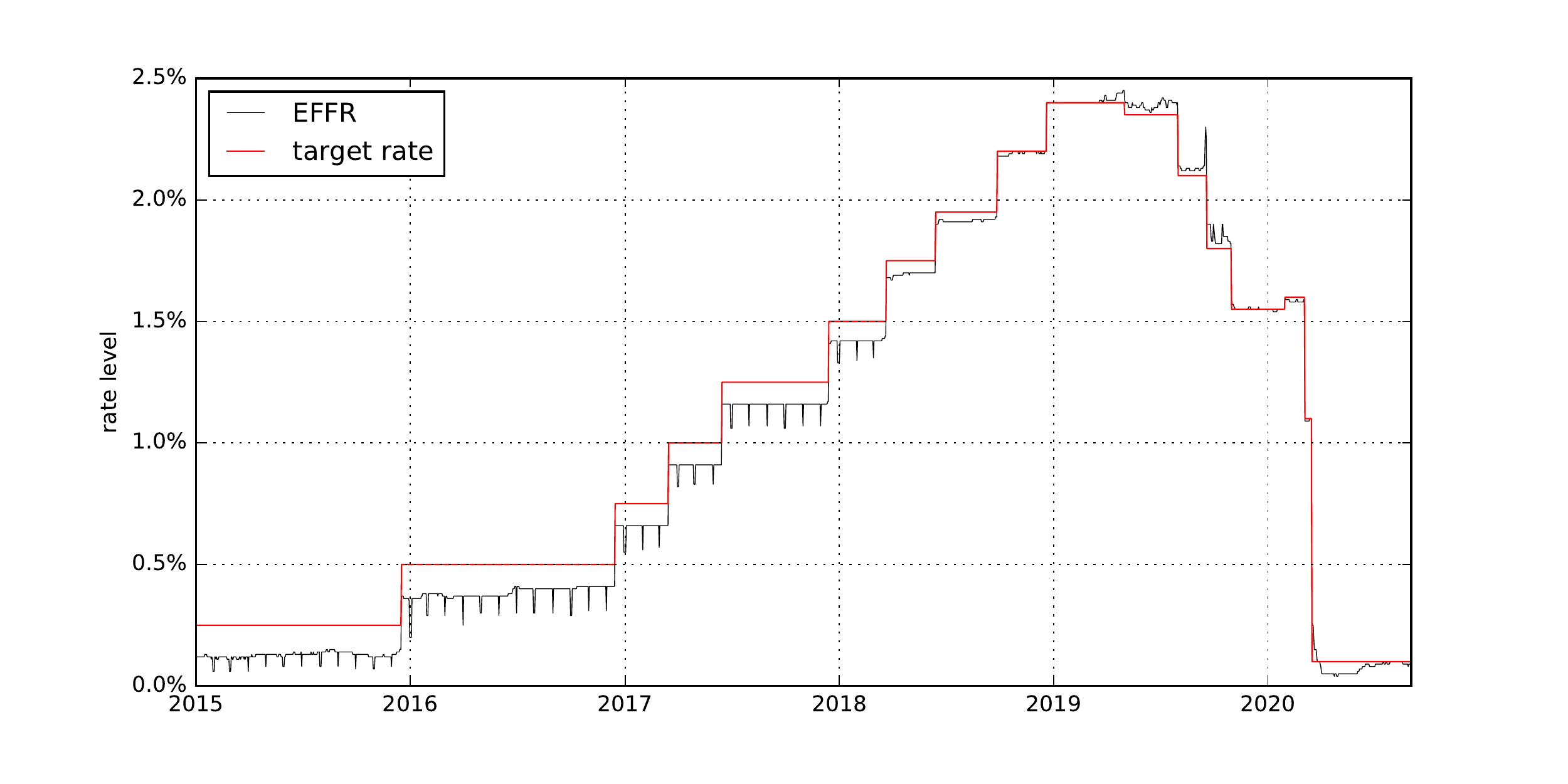}}
\caption{EFFR and FOMC target rate history}\label{target_effr}
\end{figure}

\begin{figure}[t]
\centerline{\includegraphics[scale=0.65]{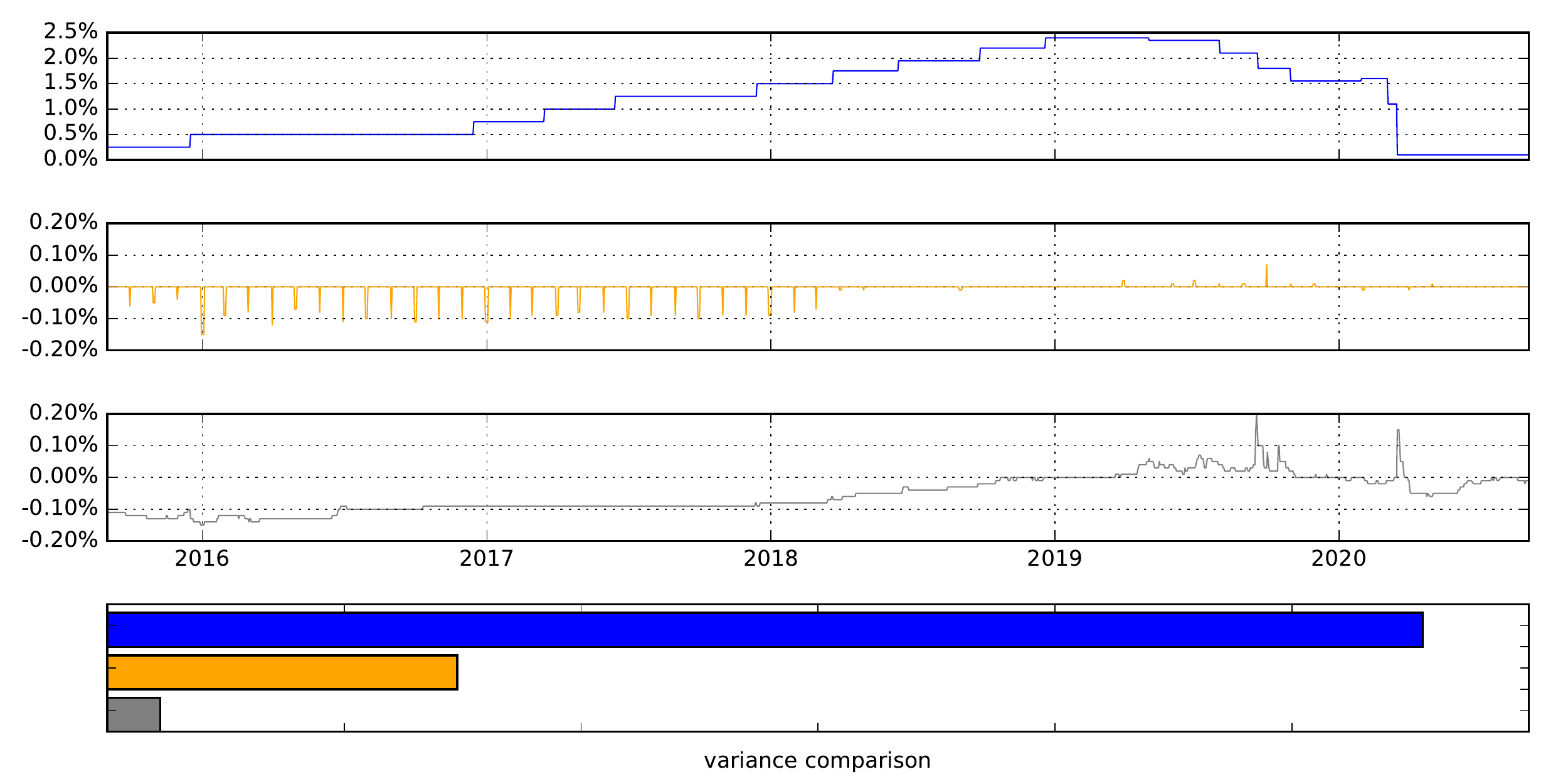}}
\caption{EFFR breakdown in vertical order (i) target rates (ii) end--of--month spikes (iii) residual (iv) variance contribution}\label{effr_breakdown}
\end{figure}

In this section, the EFFR is examined by breaking it down into distinct components. A comparison of EFFR and the Fed Funds target rate since the beginning of 2015, see Fig. \ref{target_effr}, demonstrates the low volatility in deviations of EFFR from the target rate. The target rate therefore must be a major component of the EFFR dynamics. Another feature of Fed Funds empirical data in the earlier part of the of the five years covered by Fig. \ref{target_effr} are end--of--month downward spikes. These spikes used to occur as a result of certain regulations prescribing the last day of the month as a measurement day for reporting regulatory capital, resulting in a temporary imbalance in the demand--supply for excess reserve funds.
It is instructive to deconstruct the EFFR $r_E(t)$ rate into the two components, discontinuous at known times, and a residual such that:
\begin{gather}
r_E(t)=r_P(t)+\Delta{}r_Z(t)+\zeta(t)
\end{gather}
The first component $r_P$, the policy target rate is directly observable as the IOER rate. The second component, $\Delta{}r_Z$ the end--of--month spike, can be deduced from the data. Here we place any changes to the rate on the last trading day of the month regardless of magnitude, a sufficient approach for the qualitative analysis in this section. The third component $\zeta$ captures any residual noise in addition to the first two components. The variance of the daily changes in each component, shown in Fig. \ref{effr_breakdown}, is an indicator of the relative contribution of each component. It is clear that over the 5 years of data used to produce these results, the target rate is the main factor in EFFR dynamics, followed by end--of--month spikes, with only a small contribution from the residual.

\begin{figure}[t]
\centerline{\includegraphics[scale=0.65]{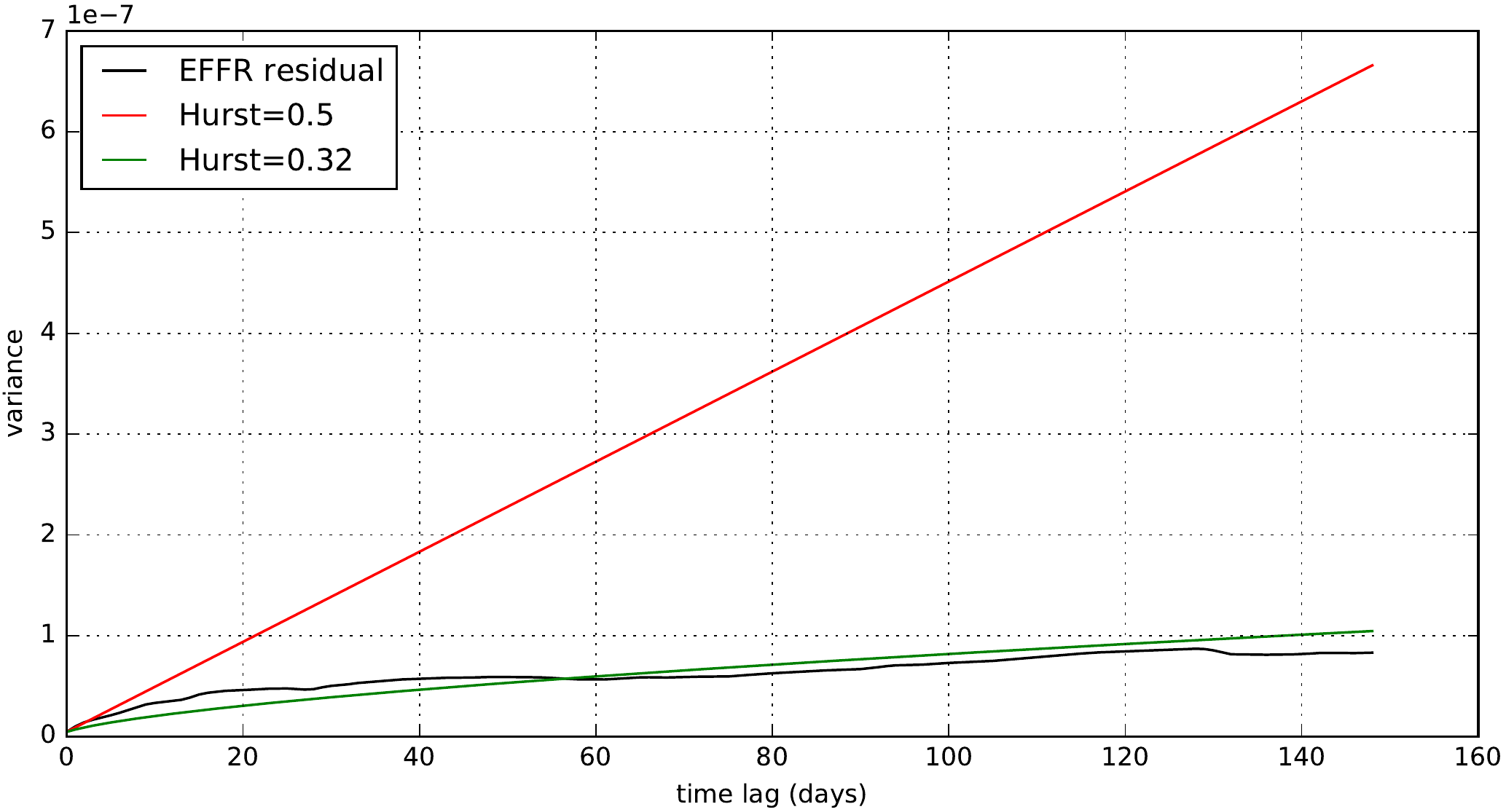}}
\caption{Variance for the difference of the residual noise time series for different lags, compared to theoretical Hurst exponent values}\label{effr_mr}
\end{figure}

The existence of mean reversion in the residual is examined by finding an approximate Hurst exponent $h$ for the time series of $\zeta$. The Hurst exponent relates the variance of the lagged difference to the lag size as follows:
\begin{gather}
\text{Var}\big[\zeta(t+\tau)-\zeta(t)\big]\sim{}\tau^{2h}
\end{gather}
A Hurst exponent value of 0.5 indicates a Brownian motion, $h<0.5$ indicates presence of mean reversion. For the residual noise time series we estimate $h=0.31$, see Fig. \ref{effr_mr}, indicating mean reversion.

In summary, the empirical characteristics of EFFR break down into the following components: piecewise flat target rates, followed by spikes occurring on known days and mean reverting residual noise. The correlation between the three components is close to zero, with the exception of a slightly negative correlation between the residual and the target rates. The negative correlation is due to a small lag between target rate changes and EFFR adapting the full magnitude of the change, temporarily changing the spread in the opposite direction to the target rate change.

\subsection{Secured Overnight Funding Rate}

Shortly following the well--publicised LIBOR manipulation scandals, the Financial Stability Board and Financial Stability Oversight Council highlighted one of the key problems related to the reference rate to be the decline in transactions underpinning LIBOR and the associated structural risks to the financial system.\footnote{See \citeasnoun{arrc2}, page 1.} As argued in \citeasnoun{Schrimpf2019}, partly to blame for the decline in interbank term lending are the inflated excess reserves discussed in the previous section.\footnote{This suggests an interesting causal link between the financial crisis, the Federal Reserve response and the emergence of SOFR by linking the decline in LIBOR transactions to excess reserves.} In response, the Federal Reserve convened the Alternative Reference Rates Committee (ARRC)\footnote{See https://www.newyorkfed.org/arrc} to explore alternative reference rates. In June 2017, the ARRC formally announced the Secured Overnight Financing Rate (SOFR) as the replacement for LIBOR. A key criterion for the choice was the large volume of transactions behind SOFR, translating to it being more representative of bank's funding costs and less susceptible to manipulation. The calculation of SOFR is based on overnight repo transactions, which in 2017 averaged around \$700b in daily transactions\footnote{For details see \citeasnoun{arrc2}, page 7.} (compared to less than \$1b for US dollar LIBOR).

\begin{figure}[t]
\centerline{\includegraphics[scale=0.65]{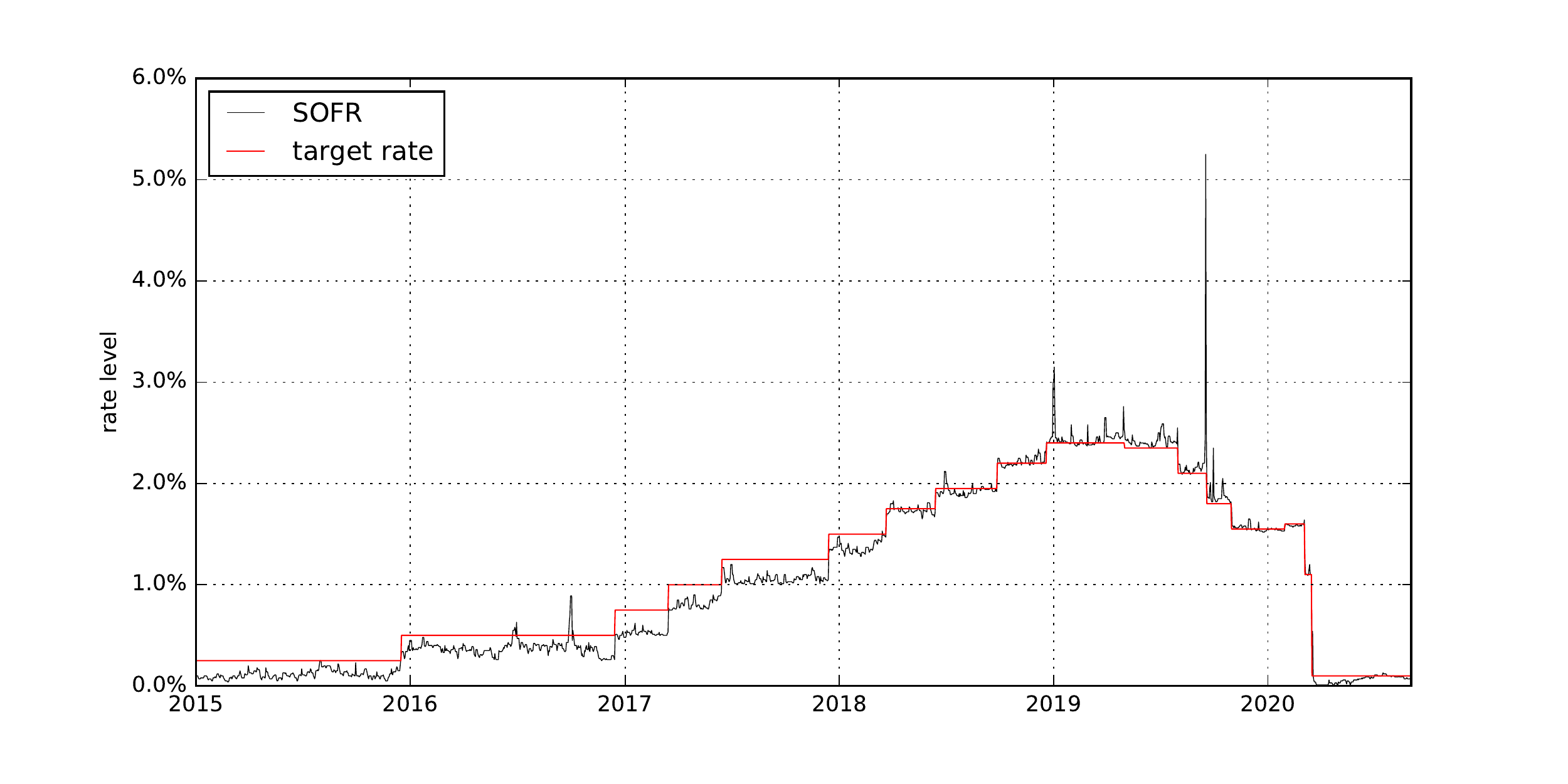}}
\caption{SOFR and FOMC target rate history}\label{target_sofr}
\end{figure}

\begin{figure}[t]
\centerline{\includegraphics[scale=0.65]{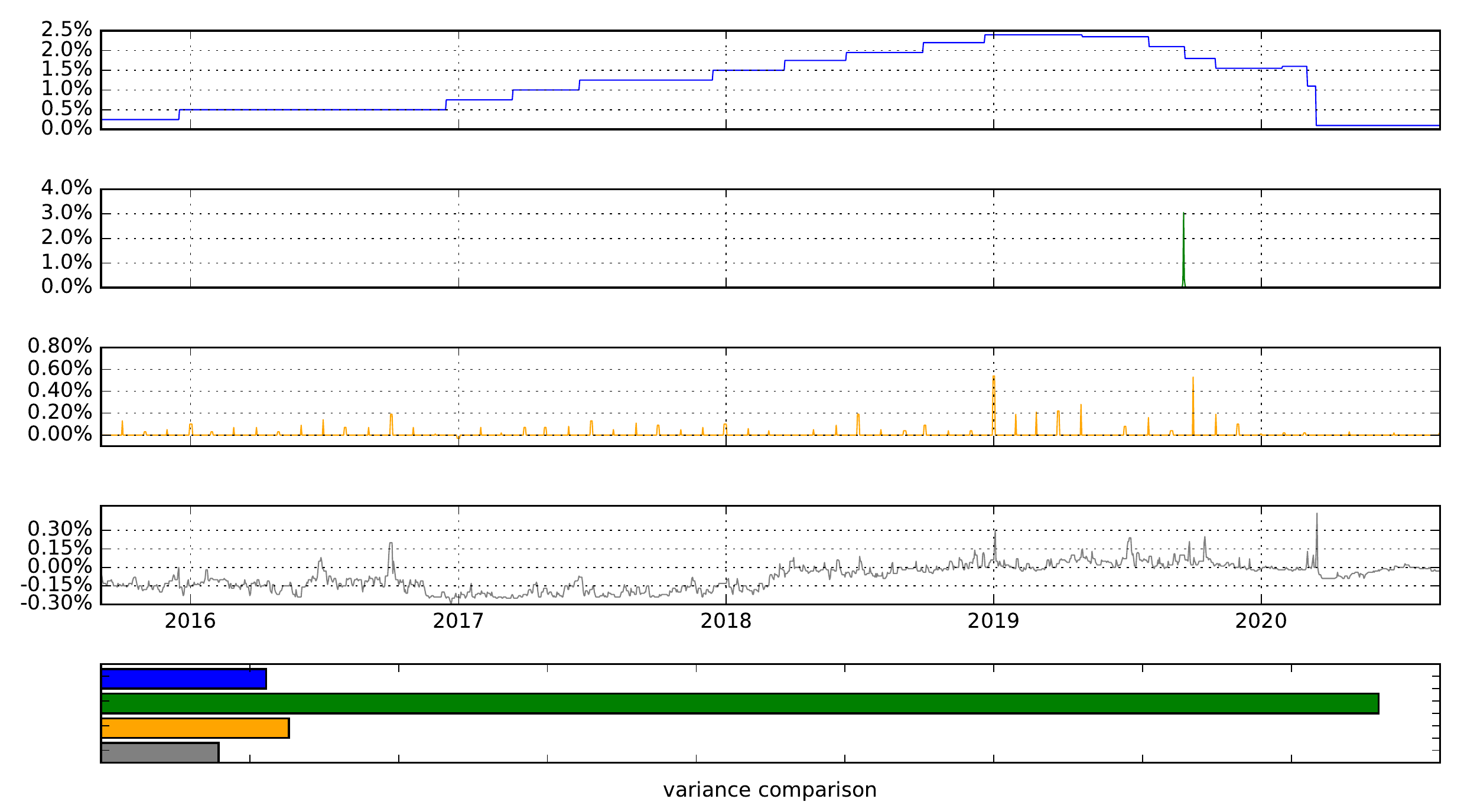}}
\caption{SOFR breakdown in vertical order (i) target rates (iii) non-eom spike (iii) end--of--month spikes (iv) residual (v) variance contribution}\label{sofr_breakdown}
\end{figure}

Official SOFR fixings have been calculated as far back 2014 and can be seen in comparison to the target rate in Fig. \ref{target_sofr}. Three features stand out, firstly SOFR appears to follow a stepwise function, suggesting that similarly to EFFR the Fed Funds target rate plays an important role in the SOFR dynamic. Another aspect is that SOFR is substantially more volatile than EFFR. A third feature is the prominence of spikes, most of which, similarly to EFFR, occur on the last trading day of the month. The end--of--month spikes are related to the measurement of dealers' balance sheet exposures at month--end for regulatory purposes. This single snapshot approach incentivises the management of exposures around reporting dates, which as explained in \citeasnoun{Schrimpf2019} has been resulting in increases in the SOFR rate on end--of--month dates. Therefore the main components of SOFR can be characterised as follows, see Fig. \ref{sofr_breakdown}:

\begin{gather}
r_S(t)=r_P(t)+\Delta{}r_Z(t)+\Delta{}r_J(t)+\zeta_s(t)
\end{gather}

Here $r_S(t)$ is the SOFR observation at time $t$, $r_P(t)$ the policy target rate, $\Delta{}r_Z(t)$ the end--of--month SOFR spikes, $\Delta{}r_J(t)$ spikes not occurring on end--of--month dates and $\zeta_s(t)$ the residual. The spikes not occurring on the last day of the trading month are the most prominent in terms of contribution to the net variance over the period, however this is due to only one very large spike occurring in September 2019. This particular spike occurred on a day of large corporate tax payments and Treasury bond expiries, therefore it could be argued that the date was predictable. The next largest contribution comes from the end--of--month spike component, followed closely by the policy target rate component.

\begin{figure}[t]
\centerline{\includegraphics[scale=0.65]{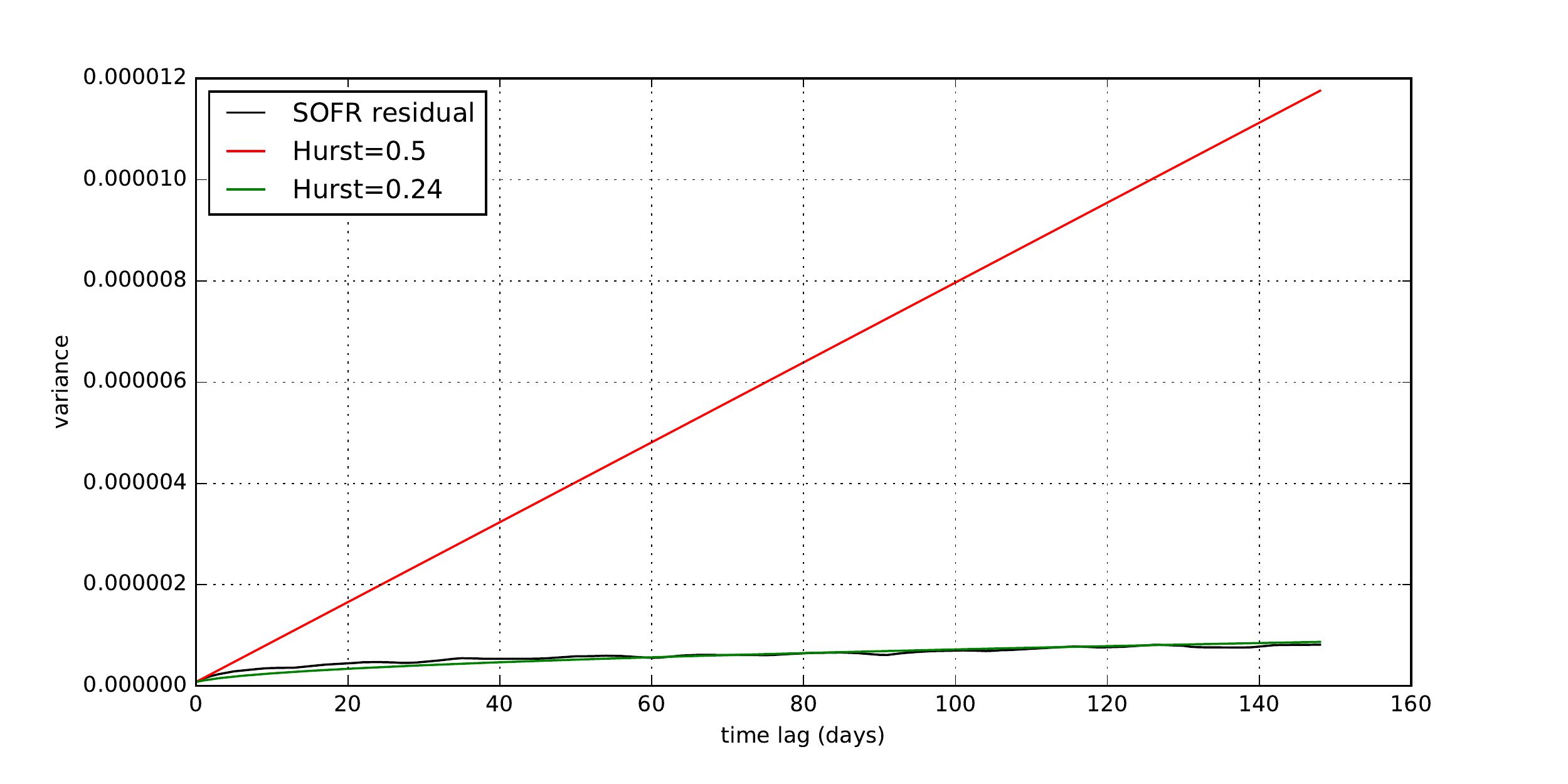}}
\caption{Variance for the difference of SOFR residual time series for different lags, compared to theoretical Hurst exponent values}\label{sofr_mr}
\end{figure}

In contrast to EFFR, the contribution from the residual component is in the same order of magnitude as the target rate component as well as the end--of--month component. Using the same approach as for EFFR, we see that the SOFR residual also exhibits strong mean reversion with an estimated Hurst parameter $h=0.24$, see Fig. \ref{sofr_mr}. In summary, the components of SOFR mostly mirror the components of EFFR, but with different contributions to the overall variance.

\section{Modelling Short Rates With Discontinuities At Known Times}\label{sec:model}
To reflect the empirical features outlined in the previous section, we assume a three--component model driven by independent factors and construct it within the HJM framework. The three components include a step component to reflect the target rate dynamics, a spike component for spikes occurring at known times and a continuous diffusion component for the residual noise. Define a set of independent Brownian motions $W$ comprising of subsets of Brownian motions $W^P$, $W^Z$, $W^V$ related to the step, spike and continuous components respectively, where $W^P=[W_1^P,...,W_m^P]$, $W^Z=[W_1^Z,...,W_n^Z]$ and where $W=[W_1,...,W_{m+n+1}]=[W_1^P,...,W_m^P,W_1^Z,...,W_n^Z,W^V]$. Under the spot risk--neutral measure, we have in the HJM framework:
\begin{gather}
f(t,T) = f(0,T)+\sum\limits_{i=1}^{m+n+1}\int\limits_0^t\sigma_i(u,T)\int\limits_u^T\sigma_i(u,s)dsdu+\sum\limits_{i=1}^{m+n+1}\int\limits_0^t\sigma_i(s,T)dW_i(s)
\end{gather}
Define:
\begin{gather}
	\sigma_i(t,T)=\mathbbm{1}(i\leq{}m)\sigma^P_i(t,T)+\mathbbm{1}(m<i\leq{}m+n)\sigma^Z_{i-m}(t,T)+\mathbbm{1}(i=m+n+1)\sigma^V(t,T)
\end{gather}
Therefore:
\begin{align}
\begin{split}
\sum\limits_{i=1}^{m+n+1}\int\limits_0^t\sigma_i(s,T)dW_i(s) &=\sum\limits_{i=1}^m\int\limits_0^t\sigma^P_i(s,T)dW^P_i(s)\\
&+\sum\limits_{i=1}^n\int\limits_0^t\sigma^Z_i(s,T)dW^Z_i(s)\\
&+\int\limits_0^t\sigma^V(s,T)dW^V(s)
\end{split}
\end{align}
and
\begin{align}
\begin{split}
	\sum\limits_{i=1}^{m+n+1}\int\limits_0^t\sigma_i(u,T)\int\limits_u^T\sigma_i(u,s)dsdu&=\sum\limits_{i=1}^m\int\limits_0^t\sigma^P_i(u,T)\int\limits_u^T\sigma^P_i(u,s)dsdu\\&+\sum\limits_{i=1}^n\int\limits_0^t\sigma^Z_i(u,T)\int\limits_u^T\sigma^Z_i(u,s)dsdu\\&+\int\limits_0^t\sigma^V(u,T)\int\limits_u^T\sigma^V(u,s)dsdu
\end{split}
\end{align}
therefore we have
\begin{gather}
f(t,T)=f^P(t,T)+f^Z(t,T)+f^V(t,T)
\end{gather}
where
\begin{gather}
f^P(t,T)=f^P(0,T)+\sum\limits_{i=1}^m\int\limits_0^t\sigma^P_i(u,T)\int\limits_u^T\sigma^P_i(u,s)dsdu+\sum\limits_{i=1}^m\int\limits_0^t\sigma^P_i(s,T)dW^P_i(s)
\end{gather}
\begin{gather}
f^Z(t,T)=f^Z(0,T)+\sum\limits_{i=1}^n\int\limits_0^t\sigma^Z_i(u,T)\int\limits_u^T\sigma^Z_i(u,s)dsdu+\sum\limits_{i=1}^n\int\limits_0^t\sigma^Z_i(s,T)dW^Z_i(s)
\end{gather}
\begin{gather}
f^V(t,T)=f^V(0,T)+\int\limits_0^t\sigma^V(u,T)\int\limits_u^T\sigma^V(u,s)dsdu+\int\limits_0^t\sigma^V(s,T)dW^V(s)
\end{gather}
similarly for the short rate:
\begin{gather}
r(t)=r^P(t)+r^Z(t)+r^V(t)
\end{gather}
and zero coupon bonds:
\begin{gather}
B(t,T)=B^P(t,T)B^Z(t,T)B^V(t,T)
\end{gather}
We now proceed to discuss the modelling of each component in more detail.

\subsection{Target Rate Step Model}\label{sec:target_rate_model}
The main empirical feature of the target rate is that it is piecewise flat between the FOMC meeting dates at which a policy change has occurred. Most of the meetings are scheduled at least one year ahead of time with the exception of emergency meetings.\footnote{Since 2015 there have been 47 meetings (including 3 emergency meetings), of which 17 resulted in a target rate change}

\begin{figure}[t]
\centerline{\includegraphics[scale=0.65]{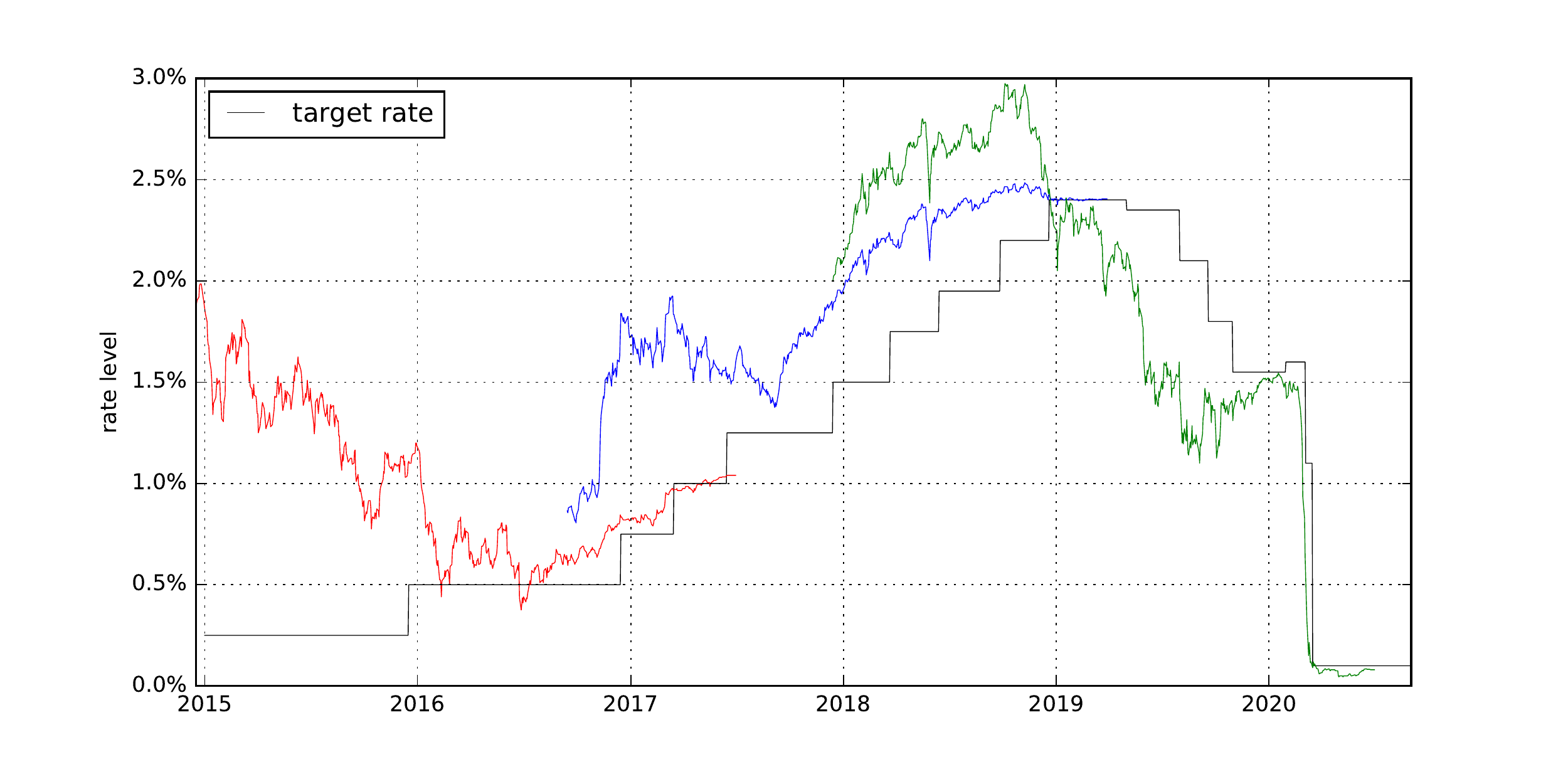}}
\caption{Target rate and various forward rates implied by specific 30--day Fed Funds futures}\label{target_rate}
\end{figure}
Forward target rates do not trade directly, however the nature of their dynamics can be deduced from 30-day Fed Fund Futures which trade on the closely related EFFR rate. Fig. \ref{target_rate} shows the historical target rate and various forward rates implied from specific futures contracts. The point at which the forward rates end and meet the target rate coincides with the expiry of the futures contracts.\footnote{Futures without an FOMC date in the reference month were chosen such that the target rate is expected to be flat over the contract month and therefore the price of the futures reflects the expected target rate for that month plus a spread rather than reflecting two flat periods before and after the FOMC date.} In contrast to the target rate, the dynamics of target forward rates are more diffusive and do not jump at deterministic dates. Jumps conceivably could occur on unexpected dates, reflecting sudden large changes in market sentiment, but in this paper we focus only on the diffusive aspect of forward rates. This is the main contribution of the present paper: Having observed that empirically the short rate (EFFR or SOFR) follows dynamics determined primarily by jumps at known times, but forward rates follow primarily diffusive dynamics, we construct a model which reconciles these two (naively contradictory) observations.

An interpretation of the forward target rates is that they reflect the expectations of prospective FOMC target rate changes. The diffusive dynamics of forward rates then reflects the nature of the changes in those expectations. From this perspective, the expectations corresponding to each scheduled FOMC meetings are not independent of each other. In some circumstances, for example, a change in the overall Federal Reserve monetary policy stance, they will be positively correlated. In other cases, where for example the aggregated change to the target rate over some period of time is anticipated but the timing is less certain, the expectations may be negatively correlated to each other as the expected timing but not the net outcome evolves.

Therefore the target rate model is motivated by the following empirical features. The target rate represented by the short rate $r^P(t)$ must be piecewise flat with respect to $t$. The forward rate with maturity $T$ evolves diffusively with respect to $t$ until the FOMC meeting immediately preceding maturity $T$, reflecting the expectations of any FOMC policy target rate change. We construct a model which reconciles these features, reflecting both the discontinuous nature of the short rate and diffusively evolving forward rates.

\subsubsection{Forward Rates}
We construct the model for forward rates such that they are driven by the evolution of expectations associated with FOMC target rate changes, where the target rate change for each scheduled meeting date evolves under its own dynamic. Define the forward rate $f(t,T)$ dynamics under the empirical measure as follows:
\begin{gather}\label{eq:emp_forward}
	f^P(t,T)=f^P(0,T)+\alpha(t,T)+\sum\limits_{i=1}^n\int\limits_0^t\xi_i(s,T)dZ_i(s)
\end{gather}
where $f^P(0,T)$ is the initial term structure of forward rates, $\alpha(t,T)$ a deterministic drift and $dZ_i(s)$ the Wiener increment corresponding to the $i_{th}$ FOMC date with correlation $dZ_i(t)dZ_j(t)=\rho_{i,j}dt$. The stochastic term is defined as follows:
\begin{gather}\label{eq:corr_sigma}
	\xi_i(t,T)=\xi_i\mathbbm{1}(t<x_i)\mathbbm{1}(T\geq{}x_i)
\end{gather}
where $x_i$ denotes the $i^{th}$ FOMC meeting date. The intuition behind this construction is that each stochastic factor corresponds to an FOMC date and any changes to the target rate are carried forward from that date. The indicator function $\mathbbm{1}(T\geq{}x_i)$ ensures that the $i^{th}$ factor is only applied to forwards with maturities greater or equal to $x_i$. For any maturities prior to the first meeting date $T<x_1$, therefore $\mathbbm{1}(T\geq{}x_i)=0,\forall{}i\geq{}1$, thus ensuring no diffusion for forward rates with maturities prior to the first FOMC meeting date. The indicator function $\mathbbm{1}(t<x_i)$ terminates the diffusion from the $i^{th}$ stochastic factor on the corresponding FOMC date. Solving the integral, see \eqref{appendix_A1}, yields:
\begin{gather}
f^P(t,T)=f^P(0,T)+\alpha(t,T)+\sum\limits_{i=1}^n\xi_i\mathbbm{1}(T\geq{}x_i)Z_i(t\wedge{}x_i)
\end{gather}
To demonstrate the behaviour of the model with an example, let $x_2<T<x_3$ and $t<x_1$:
\begin{gather}
f^P(t,T)=f^P(0,T)+\alpha(t,T)+\xi_1Z_1(t)+\xi_2Z_2(t)
\end{gather}
The interpretation being that both stochastic factor corresponding to FOMC dates $x_1$ and $x_2$ impact the forward rate up to time $t$. Any stochastic factors beyond $x_2$ do not apply since the forward rate matures prior to $x_3$. Now let $x_1<t<x_2$:
\begin{gather}
f^P(t,T)=f^P(0,T)+\alpha(t,T)+\xi_1Z_1(x_1)+\xi_2Z_2(t)
\end{gather}
In this case the first stochastic factor terminates at $x_1$, prior to $t$. That is the expectations of the target rate change at time $x_1$ evolve diffusively only up until this date.
\subsubsection{Short Rates}
The forward rate dynamics are constructed to create the piecewise dynamic in the short rate, which can be derived from \eqref{eq:emp_forward} by setting $r(t)=f(t,t)$:
\begin{gather}
	r^P(t)=f^P(t,t)=f^P(0,t)+\alpha(t,t)+\sum\limits_{i=1}^n\int\limits_0^t\xi_i(s,t)dZ_i(s)
\end{gather}
solving the integral, see \eqref{appendix_A2}, yields:
\begin{gather}
	r^P(t)=f^P(0,t)+\alpha(t,t)+\sum\limits_{i=1}^n\xi_i\mathbbm{1}(t\geq{}x_i)Z_i(x_i)
\end{gather}
From which it is evident that the short rate has no diffusion up until the first FOMC date at which point it picks up all the diffusion from the forward rate accumulated up until this point in time. To see this, for $t<x_1$ we have:
\begin{gather}
	r^P(t)=f^P(0,t)+\alpha(t,t)
\end{gather}
for $x_1<t<x_2$ we have:
\begin{gather}
	r^P(t)=f^P(0,t)+\alpha(t,t)+\xi_1Z_1(x_1)
\end{gather}
for $x_2<t<x_3$:
\begin{gather}
	r^P(t)=f^P(0,t)+\alpha(t,t)+\xi_1Z_1(x_1)+\xi_2Z_2(x_2)
\end{gather}
In general, the accumulated diffusion for the forward rates creates discontinuities in the short rate on FOMC dates, reflecting the empirical behaviour for the target rate and the associated forward rates.
\subsubsection{Decomposition to Independent Factors}
The model can be easily transformed to independent factors which will make it consistent with the HJM framework, thus allowing derivation of risk--neutral dynamics. Define $\Sigma$ to be the covariance matrix of the vector $dZ=[dZ_1,...,dZ_n]$. To transform the system to independent factors we seek to find a transformation matrix $\lambda$, such that $\Sigma=\lambda\lambda^T$ which is applied using $dZ=\lambda{}dW^P$, to result in a vector of uncorrelated Wiener increments $dW^P=[dW^P_1,...,dW^P_n]$. Therefore:
\begin{gather}\label{eq:transform}
dZ_i=\sum\limits_{j=1}^n\lambda_{i,j}dW^P_j
\end{gather}
we can rewrite the forward rate dynamics with respect to uncorrelated factors:
\begin{align}
\begin{split}
\sum\limits_{i=1}^n\int\limits_0^t\xi_i(s,T)dZ_i(s)&=\sum\limits_{i=1}^n\int\limits_0^t\sigma_i\mathbbm{1}(s<x_i)\mathbbm{1}(T\geq{}x_i)dZ_i(s)\\
&=\sum\limits_{i=1}^n\int\limits_0^t\xi_i\mathbbm{1}(s<x_i)\mathbbm{1}(T\geq{}x_i)\sum\limits_{j=1}^n\lambda_{i,j}dW^P_j(s)\\
&=\sum\limits_{j=1}^n\int\limits_0^t\sigma^P_j(s,T)dW^P_j(s)
\end{split}
\end{align}
where
\begin{gather}
\sigma^P_j(t,T)=\sum\limits_{i=1}^n\xi_i\lambda_{i,j}\mathbbm{1}(t<x_i)\mathbbm{1}(T\geq{}x_i)
\end{gather}
\subsubsection{Forward Rates Under the Spot Risk--Neutral Measure}
We can now formulate the risk neutral dynamics by using the result from HJM. Under the spot risk--neutral measure we have:
\begin{gather}
	f^P(t,T)=f^P(0,T)+\sum\limits_{j=1}^n\int\limits_0^t\sigma^P_j(u,T)\int\limits_u^T\sigma^P_j(u,s)dsdu+\sum\limits_{j=1}^n\int\limits_0^t\sigma^P_j(s,T)dW^P_j(s)
\end{gather}
Therefore, see \eqref{appendix_A3} and \eqref{appendix_A5}, we get:
\begin{align}
\begin{split}
f^P(t,T)=f^P(0,T)&+\sum\limits_{j=1}^n\sum\limits_{q=1}^n\sum\limits_{i=1}^n\xi_q\xi_i\lambda_{q,j}\lambda_{i,j}\mathbbm{1}(T\geq{}x_{q\vee{}i})(T-x_i)[t\wedge{}x_q\wedge{}x_i]\\
&+\sum\limits_{j=1}^n\sum\limits_{i=1}^n\xi_i\lambda_{i,j}\mathbbm{1}(T\geq{}x_i)W^P_j(t\wedge{}x_i)\label{fPdyn}
\end{split}
\end{align}

\subsubsection{Short Rates Under the Spot Risk--Neutral Measure}
Short rate dynamics can are obtained as follows:
\begin{gather}
r^P(t) = f^P(0,t)+\sum\limits_{j=1}^n\int\limits_0^t\sigma^P_j(u,t)\int\limits_u^t\sigma^P_j(u,s)dsdu + \sum\limits_{j=1}^n\int\limits_0^t\sigma^P_j(s,t)dW^P_j(s)
\end{gather}
Therefore, see \eqref{appendix_A6} and \eqref{appendix_A8}, we get:
\begin{align}
\begin{split}
r^P(t) &= \underbrace{f^P(0,t)+\sum\limits_{j=1}^n\sum\limits_{q=1}^n \sum\limits_{i=1}^n\xi_q\xi_i\lambda_{q,j}\lambda_{i,j}\mathbbm{1}(t\geq{}x_{q\vee{}i})(t-x_i)[x_q\wedge{}x_i]}_{\text{deterministic term (**)}}\\
&\qquad + \underbrace{\sum\limits_{j=1}^n\sum\limits_{i=1}^n\xi_i\lambda_{i,j}\mathbbm{1}(t\geq{}x_i)W^P_j(x_i)}_{\text{stochastic term (*)}}
\label{rPdyn}
\end{split}
\end{align}
The stochastic term (*) follows piecewise constant dynamics, jumping almost surely at each $x_i$.\footnote{At this point, one might object that in reality, rates do not jump at every FOMC meeting date. However, one could argue that this is because target rates are only updated in discrete increments. Our model could be extended to reflect this, but as a first approximation, we'll accept the implication of a continuous distribution of jump sizes for now, with jumps occurring at every FOMC meeting date.} Because at present we are only modelling the target rate, we would want the paths of $r^P(t)$ to be constant between FOMC meeting dates. This implies that the deterministic term (**) should not depend on $t$, i.e., the dependence on $t$ of the triple sum must cancel against the dependence on $t$ of the initial term structure $f^P(0,t)$.\footnote{Note that the term $(t-x_i)$ appearing in the triple sum reflects a feature of a classical Gaussian term structure model without mean reversion (as noted, for example, in \citeasnoun{Sch&Som:98}), that the term structure of forward rates endogenously steepens ever more (see also (\ref{fPdyn}) above) as time passes --- this can be avoided by introducing mean reversion.} When considering a time horizon of two years or less (as we do in the empirical section of this paper), the triple sum in (**) is practically flat in $t$, so this is consistent with an initial term structure $f^P(0,t)$ which is approximately constant between FOMC meeting dates. Note, however, that (\ref{rPdyn}) implies that if we require the paths of $r^P(t)$ to be constant between FOMC meeting dates, we cannot arbitrarily choose an interpolation method for the initial term structure. In particular, requiring piecewise constant paths of $r^P(t)$ precludes applying the popular Nelson/Siegel interpolation to the initial term structure.\footnote{\citeasnoun{Skov2020} show that a three--factor Gaussian arbitrage--free Nelson/Siegel model is well suited for the SOFR futures market, but they do not include the time series of SOFR itself in their estimation, i.e., their objective is not to match the SOFR dynamics, which have a substantial piecewise flat component.}

\subsubsection{Bond Prices}
Bond prices can be written as follows:
\begin{gather}\label{eq:target_bond_price}
	B^P(t,T)=\text{exp}\bigg(-\int\limits_t^Tf^P(t,s)ds\bigg)=\frac{B^P(0,T)}{B^P(0,t)}\text{exp}\bigg(a(t,T)+b(t,T)\bigg)
\end{gather}
where
\begin{align}
\begin{split}
a(t,T)&=-\int\limits_t^T\sum\limits_{j=1}^n\sum\limits_{q=1}^n\sum\limits_{i=1}^n\xi_q\xi_i\lambda_{q,j}\lambda_{i,j}\mathbbm{1}(s\geq{}x_{q\vee{}i})(s-x_i)[t\wedge{}x_q\wedge{}x_i]ds\\
&=-\sum\limits_{j=1}^n\sum\limits_{q=1}^n\sum\limits_{i=1}^n\xi_q\xi_i\lambda_{q,j}\lambda_{i,j}[t\wedge{}x_q\wedge{}x_i]\int\limits_t^T\mathbbm{1}(s\geq{}x_{q\vee{}i})(s-x_i)ds\\
&=-\sum\limits_{j=1}^n\sum\limits_{q=1}^n\sum\limits_{i=1}^n\xi_q\xi_i\lambda_{q,j}\lambda_{i,j}[t\wedge{}x_q\wedge{}x_i][I_1-I_2]
\end{split}
\end{align}
where
\begin{align}
\begin{split}
I_1 &= \int\limits_0^T\mathbbm{1}(s\geq{}x_{q\vee{}i})(s-x_i)ds\\
&= \int\limits_{x_{q\vee{}i}}^T(s-x_i)ds=\mathbbm{1}(T\geq{}x_{q\vee{}i})[T(\frac{T}{2}-x_i)-x_{q\vee{}i}(\frac{x_{q\vee{}i}}{2}-x_i)]
\end{split}
\end{align}
\begin{gather}
I_2=\int\limits_0^t\mathbbm{1}(s\geq{}x_{q\vee{}i})(s-x_i)ds=\mathbbm{1}(t\geq{}x_{q\vee{}i})[t(\frac{t}{2}-x_i)-x_{q\vee{}i}(\frac{x_{q\vee{}i}}{2}-x_i)]
\end{gather}

\begin{align}
\begin{split}
b(t,T)=&-\int\limits_t^T\sum\limits_{j=1}^n\sum\limits_{i=1}^n\xi_i\lambda_{i,j}\mathbbm{1}(s\geq{}x_i)W^P_j(t\wedge{}x_i)ds\\
&=-\sum\limits_{j=1}^n\sum\limits_{i=1}^n\xi_i\lambda_{i,j}W^P_j(t\wedge{}x_i)\int\limits_t^T\mathbbm{1}(s\geq{}x_i)ds\\
&=-\sum\limits_{j=1}^n\sum\limits_{i=1}^n\xi_i\lambda_{i,j}W^P_j(t\wedge{}x_i)\mathbbm{1}(T\geq{}x_i)[T-(t\vee{}x_i)]
\end{split}
\end{align}
Note that zero coupon bond prices are exponential affine functions of the $W_j^P(t\wedge x_i)$. However, unlike in classical Gauss/Markov HJM term structure models, here we cannot represent the entire term structure as an exponential affine function of $n$ factors.

\subsection{Known Spike Time Model}

\begin{figure}[t]
\centerline{\includegraphics[scale=0.65]{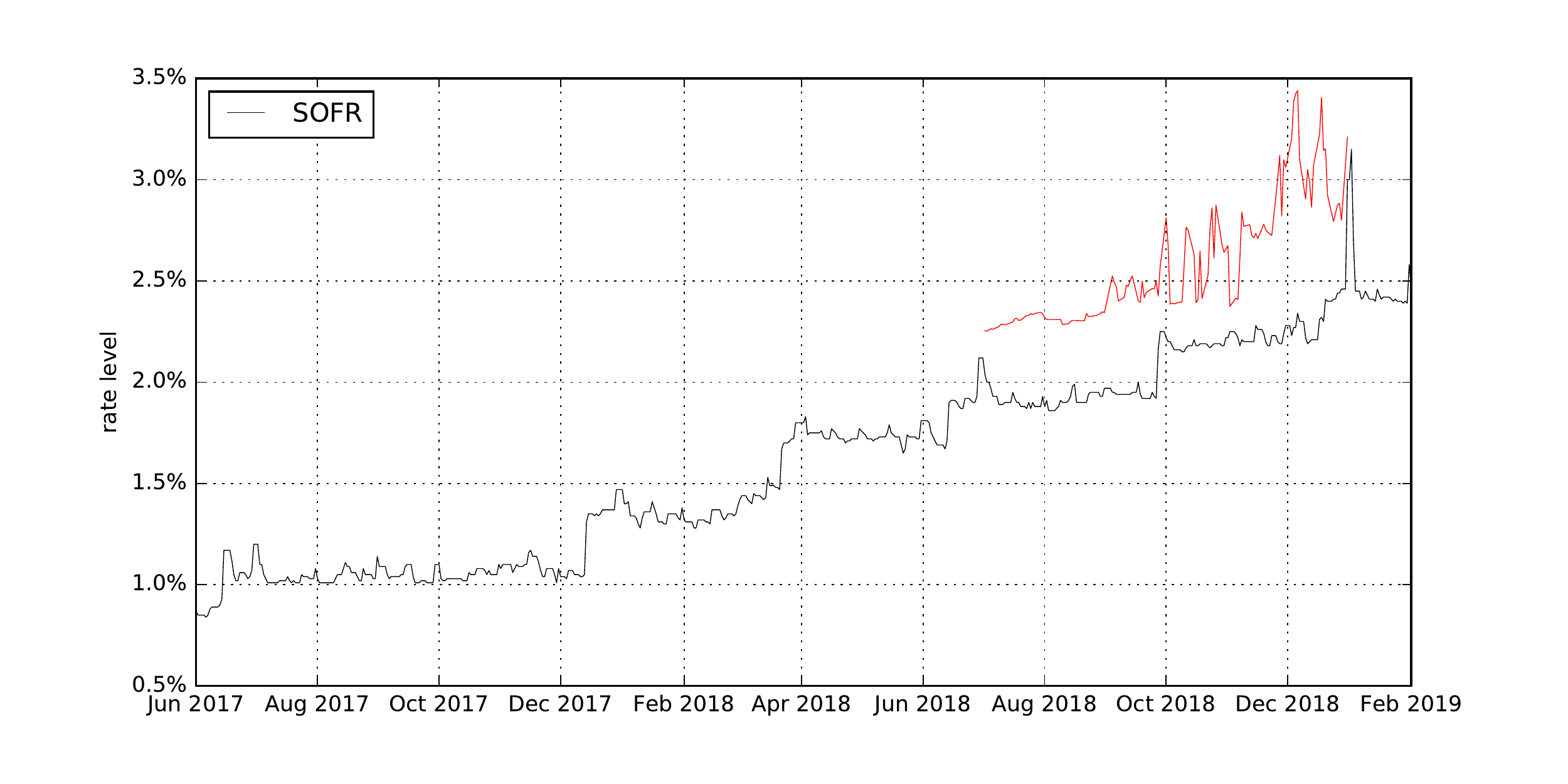}}
\caption{SOFR and a forward rate associated with the end of month December 2018 spike}\label{spike_forwards}
\end{figure}

As shown in Section \ref{sec:empirical}, spikes in the short rate are a prominent feature of EFFR and particularly SOFR dynamics. Similarly to the previous section, the forward rates associated with the spikes can be deduced from the futures market, see Fig. \ref{spike_forwards}, revealing similar empirical behaviour of forward rates evolving more diffusively rather than showing discontinuities on known dates. In this section, we adapt the approach of the previous section to reflect the occurrence of spikes at known dates.

\subsubsection{Spiked Forward Rates}
The model for forward rates is constructed such that the forward rate on each spike date $z_i$ is driven by its own independent factor. Therefore we can use the HJM result to formulate the forward rates under the risk neutral measure:
\begin{gather}
	f^Z(t,T)=f^Z(0,T)+\sum\limits_{i=1}^n\int\limits_0^t\sigma^Z_i(u,T)\int\limits_u^T\sigma^Z_i(u,s)dsdu+\sum\limits_{i=1}^n\int\limits_0^t\sigma^Z_i(s,T)dW^Z_i(s)
\end{gather}
We assume that when spikes occur, they impact a fixed period $h_i$ starting from time $z_i$.\footnote{Usually this period of time would be equivalent to 1 day but could be more if for example it is a SOFR rate set on a Friday, therefore applying for compounding and averaging payoff calculations over the weekend} Let $H_i=[z_i,z_i+h_i]$, the volatility function is defined as follows:
\begin{gather}
	\sigma^Z_i(t,T)=\sigma^Z_i\mathbbm{1}(t<z_i)\mathbbm{1}(T\in{}H_i)
\end{gather}
Therefore, see \eqref{appendix_A9} and \eqref{appendix_A11} :
\begin{gather}
f^Z(t,T)=f^Z(0,T)+\sum\limits_{i=1}^n\big(\sigma^Z_i\big)^2\mathbbm{1}(T\in{}H_i)(T-z_i)[t\wedge{}z_i]+\sum\limits_{i=1}^n\sigma^Z_i\mathbbm{1}(T\in{}H_i)W_i(t\wedge{}z_i)
\end{gather}
To demonstrate the behaviour of the model with an example, let $T\in{}H_2$ and $t<z_1$:
\begin{gather}
f^Z(t,T)=f^Z(0,T)+\big(\sigma^Z_2\big)^2(T-z_2)t+\sigma^Z_2W_2(t)
\end{gather}
The interpretation being that the forward rates only evolve when $T\in{}H_i$ up to the minimum of time $t$ or $z_i$, the beginning of the period $H_i$.

\subsubsection{Spiked Short Rates}
By construction short rates follow the spiked trajectory, we have:
\begin{gather}
r^Z(t) = f^Z(t,t) = f^Z(0,t)+\sum\limits_{i=1}^n\int\limits_0^t\sigma^Z_i(u,t)\int\limits_u^t\sigma^Z_i(u,s)dsdu + \sum\limits_{i=1}^n\int\limits_0^t\sigma^Z_i(s,t)dW^Z_i(s)
\end{gather}
Therefore, see \eqref{appendix_A12} and \eqref{appendix_A14} :
\begin{gather}
r^Z(t)=f^Z(0,t)+\sum\limits_{i=1}^n\big(\sigma^Z_i\big)^2\mathbbm{1}(t\in{}H_i)(t-z_i)z_i+\sum\limits_{i=1}^n\sigma^Z_i\mathbbm{1}(t\in{}H_i)W_i(z_i)
\end{gather}
From this it is evident that the short rate is deterministic until the spike interval over which a spike applies, with a magnitude which includes the associated forward rate diffusion accumulated up to the beginning of the interval. For example let $t\in{}H_2$:
\begin{gather}
r^Z(t)=f^Z(0,t)+\big(\sigma^Z_2\big)^2(t-z_2)z_2+\sigma^Z_2W_2(z_2)
\end{gather}
\subsubsection{Spiked Bond Prices}
The bond prices can be written as follows:
\begin{gather}
	B^Z(t,T)=\text{exp}\bigg(-\int\limits_t^Tf^Z(t,s)ds\bigg)=\frac{B^Z(0,T)}{B^Z(0,t)}\text{exp}\bigg(a(t,T)+b(t,T)\bigg)
\end{gather}
\begin{align}
\begin{split}
a(t,T)&=-\int\limits_t^T\sum\limits_{i=1}^n\big(\sigma^Z_i\big)^2\mathbbm{1}(s\in{}H_i)(s-z_i)[t\wedge{}z_i]ds\\
&=-\big(\sigma^Z_i\big)^2[t\wedge{}z_i]\int\limits_t^T\mathbbm{1}(s\in{}H_i)(s-z_i)ds\\
&=-\big(\sigma^Z_i\big)^2[t\wedge{}z_i][I_1-I_2]
\end{split}
\end{align}
where
\begin{gather}
I_1=\int\limits_0^T\mathbbm{1}(s\in{}H_i)(s-z_i)ds=\mathbbm{1}(T\geq{}z_i)\bigg[\frac{h_i^2}{2}\wedge{}\frac{(T-z_i)^2}{2}\bigg]
\end{gather}
and
\begin{gather}
I_2=\int\limits_0^t\mathbbm{1}(s\in{}H_i)(s-z_i)ds=\mathbbm{1}(t\geq{}z_i)\bigg[\frac{h_i^2}{2}\wedge{}\frac{(t-z_i)^2}{2}\bigg]
\end{gather}
and
\begin{align}
\begin{split}
b(t,T)&=-\int\limits_t^T\sum\limits_{i=1}^n\sigma^Z_iW_i(t\wedge{}z_i)\mathbbm{1}(s\in{}H_i)ds\\
&=-\sum\limits_{i=1}^n\sigma^Z_iW_i(t\wedge{}z_i)\int\limits_t^T\mathbbm{1}(s\in{}H_i)ds\\
&=-\sum\limits_{i=1}^n\sigma^Z_iW_i(t\wedge{}z_i)([(T-z_i)\wedge(T-t)\wedge{}h_i\wedge(z_i+h_i-t)]\vee0)
\end{split}
\end{align}

\subsection{Modelling the Diffusive Residual}
As shown in Section \ref{sec:empirical}, an empirical feature of the noise component of both EFFR and SOFR is mean reversion. Since the initial bond term structure is most naturally contained in the initial target rate term structure, the mean reverting Vasicek model should be sufficient to model the noise component of short rates. We present the model based on the results shown in \citeasnoun{Carmona2007}. The dynamics of the diffusive residual are given by:
\begin{gather}
dr^V(t)=(\theta-\beta{}r^V(t))dt+\sigma^VdW^V(t)
\end{gather}
The solution is given by:
\begin{gather}
	r^V(t)=e^{-\beta{}t}r^V(0)+(1-e^{-\beta{}t})\frac{\theta}{\beta}+\sigma{}^V\int\limits_0^te^{-\kappa(t-s)}dW^V(s)
\end{gather}
With forward rates:
\begin{gather}
	f^V(t,T)=r^V(t)e^{-\beta{}(T-t)}+\frac{\theta}{\beta}\bigg(1-e^{-\beta{}(T-t)}\bigg)-\frac{\theta^2}{2\beta^2}\bigg(1-e^{-\beta{}(T-t)}\bigg)^2
\end{gather}
The zero coupon bond price is given by:
\begin{gather}
B^V(t,T)=a(t,T)e^{b(t,T)r(0)}
\end{gather}
with
\begin{gather}
b(t,T)=-\frac{1-e^{-\beta{}(T-t)}}{\beta}
\end{gather}
and
\begin{gather}
a(t,T)=\frac{4\theta\beta-3\sigma^2}{4\beta^3}+\frac{(\sigma^V)^2-2\alpha\beta}{2\beta^2}T+\frac{(\sigma^V)^2-\alpha\beta}{\beta^3}e^{-\beta{}T}-\frac{(\sigma^V)^2}{4\beta^3}e^{-2\beta{}T}
\end{gather}

\section{Calibration to Futures Contracts}\label{sec:empirical_results}
This section presents results calibrating the model to Fed Funds and SOFR futures data. Fed Fund futures are used to calibrate the target rate term structure, which is then used as the basis for calibration to SOFR futures, from which we infer the term structure of forward rates related to end--of--month spikes. The time series of calibrated EFFR and SOFR forward rate vectors is used to examine how well the market anticipates FOMC policy target rate changes as well as end--of--month spikes. The time series of SOFR forward rates is then used to compare the forward looking SOFR term rates to LIBOR.
\subsection{30 day Fed Funds futures}\label{sec:ff_calibration}
Fed Funds futures contracts\footnote{Source: https://www.cmegroup.com/trading/interest-rates/stir/\\30-day-federal-fund\_contract\_specifications.html} are based based on the arithmetic average of the EFFR, denoted $r_E$ over the specified contract month. Define $m$ as the number of months from the current trading month ($m=0$), $\tau_{m,i}:=$ as the date corresponding to day $i$ in month $m$ with $n_m$ denoting the total days in month $m$.

Define the futures contract index for reference month $m$ at time $t$ as $\widetilde{F}_m(t)$, the value of a single contract is $\text{\$4,167}\times\widetilde{F}_m(t)$. The terminal value of the contract is determined as $\widetilde{F}_m(\tau_{m,n_m})=100-R_m$ where $R_m$ is the arithmetic average of the daily EFFR fixing during the contract month, settled on the first business day after the final fixing date.

Defining $R_m:=\frac{100}{n_m}\sum\limits_{i=1}^{n_m}r_E(\tau_{m,i})$, the terminal payoff is:
$$\widetilde{F}_m(\tau_{m,n_m})=100-R_m=100\bigg(1-\frac{1}{n_m}\sum\limits_{i=1}^{n_m}r_E(\tau_{m,i})\bigg)$$
Using the generic futures pricing theorem,\footnote{See \citeasnoun{Hunt2004}, Theorem 12.6.} the expected value at $t$ of the futures contract index $\widetilde{F}_m$ under the spot risk neutral measure is:
\begin{gather}
F_m(t)=E_t[\widetilde{F}_m(\tau_{m,n_m})]=100\bigg(1-\frac{1}{n_m}\sum\limits_{i=1}^{n_m}E_t[r_E(\tau_{m,i})]\bigg)
\end{gather}
The current futures contract continues to trade during the observation month, therefore the valuation needs to account for already observed values of $r_E$:
\begin{gather}
F_0(t)=100\bigg(1-\frac{1}{n_0}\bigg(\sum\limits_{i=1}^{n_0}\mathbf{1}_{(t>\tau_{0,i})}r_E(\tau_{0,i})+\sum\limits_{i=1}^{n_0}\mathbf{1}_{(t\leq\tau_{0,i})}E_t[r_E(\tau_{0,i})]\bigg)\bigg)
\end{gather}

\subsubsection{Calibration}
Fed Funds futures contracts are available for each calendar month approximately 3 years ahead of time. However the liquidity beyond 1 year deteriorates and therefore we limit the calibration to the first 12 contracts. The availability of contracts for each calendar month makes the Fed Funds futures particularly useful for extracting information regarding expected target rate changes, which are scheduled 8 times per year and never twice in the same month. Calibrating the expected policy target rate jumps from Fed Funds futures is performed by making the following assumptions.

Firstly the initial term structure of $f(0,T)$ is assumed to be piecewise flat between FOMC meeting dates. This aligns the initial term structure to the driving factors of the target rate model and therefore the daily changes in the calibrated $f(0,T)$ vector provides an empirical estimate for the dynamics of $f(t,T)$. To simplify the calibration, it is assumed that the impact on the drift component is negligible, particularly if the calibration is used to obtain the empirical dynamics of the forward rate based on daily increments obtained from the calibration. The spikes are a secondary component of EFFR empirical dynamics and are ignored in the calibration.\footnote{Since there are 12 futures contracts and 8 FOMC meetings it may be possible to extract information regarding expected EFFR spikes from futures data} We also calibrate a constant spread between EFFR and the target rate which is equivalent to assuming zero volatility in the Gaussian residual noise component of the model.

Observable market prices exist in the form of current bid and offer and last observed price, which reflects a trade at either the bid or the offer levels at the time of the transaction. We take the view that at any given time the true market state is at some point between the bid and offer prices. Closing prices which are recorded at the end of each day's trading session also reflect either the bid or the offer. Therefore the closing price could be either the offer, inferring that the bid is one price fluctuation below the closing price, or conversely infer that the offer is one price fluctuation above the closing price. Based on this reasoning we embed a minimum price fluctuation size tolerance to the calibration error $e_m(t)$ for month $m$:
$$e_m(t)=(|F_m(t)-\widetilde{F}_m(t)|-h_m)^+$$
Where the minimum fluctuation of the index for month $m$ as $h_m$ with $h_0=0.0025$ and $h_m=0.005$ for $m\neq0$. The error bounds result in better solution stability less subject to bid-ask fluctuations in the cross sectional and longitudinal data. The calibration is performed using a genetic algorithm approach based on the method developed in \citeasnoun{GellertSchlogl2019}.

\subsubsection{Results}
To analyse the dynamics of the stepwise model forward rates, the calibration is performed on daily data in the period from January 2015 to September 2020. Additionally, we measure the agreement between actual target rate changes and the corresponding change inferred from the initial term structure of calibrated forward rates. This demonstrates how well the futures market was able to predict target rate changes in the test period. It is also a good indicator of the ability of the model to translate futures data into a meaningful term structure of anticipated target rate changes.

\begin{figure}[t]
\centerline{\includegraphics[scale=0.65]{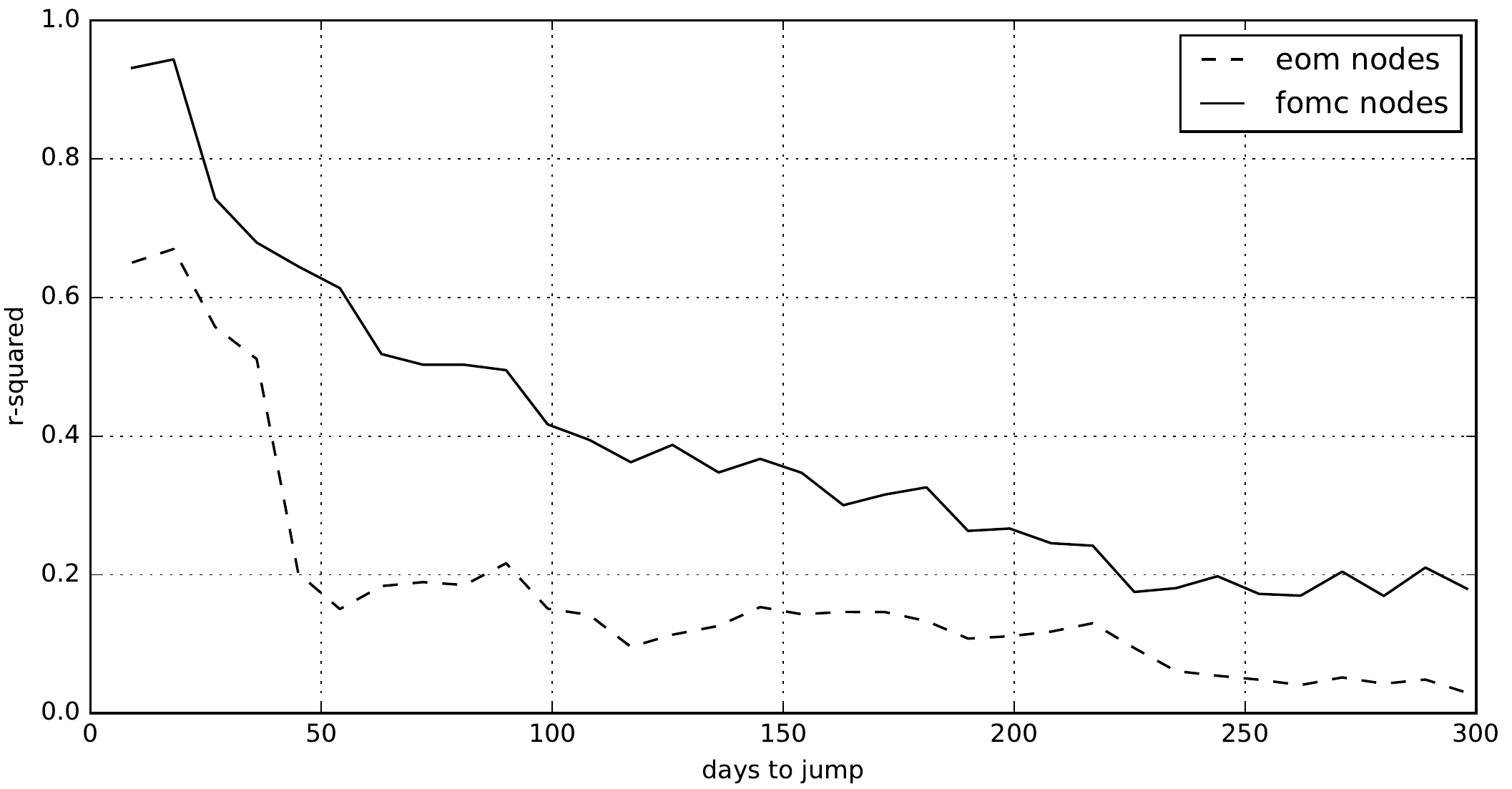}}
\caption{R-squared of EFFR realised vs forward rates for different forward periods}\label{effr_r2}
\end{figure}

To measure the agreement the R-squared is calculated between actual target rate changes $\Delta{}r^P(x_i)$ and corresponding initial forward rate term structure inferred changes $f^P(0,x_i)-f^P(0,x_i-h)$, grouped by the number of days in the forward rate term, that is the number of days between the calibration date corresponding to $t=0$ and $x_i$. The results in Fig. \ref{effr_r2} show the R-squared for increasing number of days between $x_i$ and the calibration date. For comparison, the same calculation is shown with the same piecewise flat assumption but with discontinuity dates naively set to coincide with futures contract maturities.

The results show a clear correspondence between actual and anticipated target rate changes. The correspondence deteriorates as the forward term increases but still shows evidence of some anticipation for terms over 200 days. The results comprise of a mixture of good long term anticipation of rate increases and poor anticipation of rate decreases. This can be attributed to the well communicated and regular increases in the target rate during the normalisation phase following near zero target rates. The rapid drop in target rates at the beginning of 2020 was not expected by the market, excluding this period would substantially improve the R-squared results.

\subsection{SOFR Futures}

SOFR futures are available in monthly and quarterly contract period lengths. The SOFR 1M futures contracts\footnote{Source:https://www.cmegroup.com/trading/interest-rates/stir/one-month-sofr\_contract\_specifications.html} are defined to reflect the specification of the Fed Funds 30 day futures with SOFR replacing the EFFR as the reference rate. Therefore the pricing formulas described in the previous section also apply to SOFR 1M futures.

In contrast to the monthly contracts, the final payoff of the SOFR 3M futures contracts\footnote{Source:https://www.cmegroup.com/trading/interest-rates/stir/three-month-sofr\_contract\_specifications.html} compounds SOFR, denoted by $r_s$, over IMM quarterly dates,\footnote{Third Wednesday of March, June, September and December} aligning the dates of the contracts to the LIBOR--referenced quarterly Eurodollar futures. Define $q$ as the number of IMM quarters from the current trading quarter ($q=0$), $\tau^*_{q,i}:=$ as the date corresponding to day $i$ in quarter $q$ with $n_q$ denoting the total days in quarter $q$. The SOFR 3M futures contract terminal payoff is:
$$\widetilde{F}^{s3}_q(\tau_{q,n_q})=100-R^{s3}_q$$
where $R^{s3}$ is based on SOFR compounded over the reference quarter:
$$R^{s3}=100\times\frac{360}{n_q}\bigg{[}\prod\limits_{i=1}^{n_q}\bigg{\{}\mathbf{1}_{(\tau_{q,i}\in\mathbf{b})}\bigg{(}1+\frac{d_ir_s(\tau_{q,i})}{360}\bigg{)}\bigg{\}}-1\bigg{]}$$
where $\mathbf{b}$ is the set of US government securities business days and $d_i$ the number of days the rate $r_s(\tau_{q,i})$ applies.\footnote{$d_i$ is equal to one plus the number of consecutive business days immediately following $\tau_{q,i}$}
Using the generic futures pricing theorem, the expected value at $t$ of the futures contract index $\widetilde{F}^{s3}_q$ under the spot risk neutral measure is:
\begin{gather}
F^{s3}_q(t)=E_t[\widetilde{F}^{s3}_q(\tau_{q,n_q})]=100\bigg(1-\frac{360}{n_q}\bigg{[}\prod\limits_{i=1}^{n_q}\bigg{\{}\mathbf{1}_{(\tau_{q,i}\in\mathbf{b})}\bigg{(}1+\frac{d_iE_t[r_s(\tau_{q,i})]}{360}\bigg{)}\bigg{\}}-1\bigg{]}\bigg)
\end{gather}
The current futures continues to trade during the observation quarter, therefore the valuation needs to account for already observed values of $r_s$:
\begin{gather}
F^{s3}_0(t)=E_t[\widetilde{F}^{s3}_q(\tau_{0,n_0})]=100\bigg(1-\frac{360}{n_0}\bigg{[}\prod\limits_{i=1}^{n_0}\bigg{\{}\mathbf{1}_{(\tau_{0,i}\in\mathbf{b})}\bigg{(}1+\frac{d_ir_s^*(\tau_{0,i})}{360}\bigg{\}}-1\bigg{]}\bigg)
\end{gather}
where $r_s^*(\tau_{0,i})=\mathbf{1}_{(t>\tau_{0,i})}r_s(\tau_{0,i})+\mathbf{1}_{(t\leq\tau_{0,i})}E_t[r_s(\tau_{0,i})]$

\subsubsection{Calibration}
Similarly to Fed Funds futures, SOFR 1M futures are available for each calendar month with liquidity approximately 1 year ahead, SOFR 3M futures are available between quarterly IMM dates approximately 2 years ahead of expiry. Calibrating the target rate term structure to Fed Fund futures allows the use of SOFR futures to extract information regarding the expected SOFR end--of--month spikes. We calibrate the spike component of the model to SOFR futures with similar assumptions as in the case of Fed Fund futures.

The SOFR term structure is assumed to consist of the target rate term structure obtained from Fed Fund futures, an end--of--month spike specific to the SOFR rate and a SOFR specific spread. The drift component of the spike is ignored assuming it has a negligible effect on the inferred spike forward dynamics. The spread is assumed constant for all forwards which is equivalent to the assumption of zero volatility for the noise component. The treatment related to the bid-ask spread is applied in the same way as for Fed Fund futures.

\subsubsection{Results}
\begin{figure}[t]
\centerline{\includegraphics[scale=0.65]{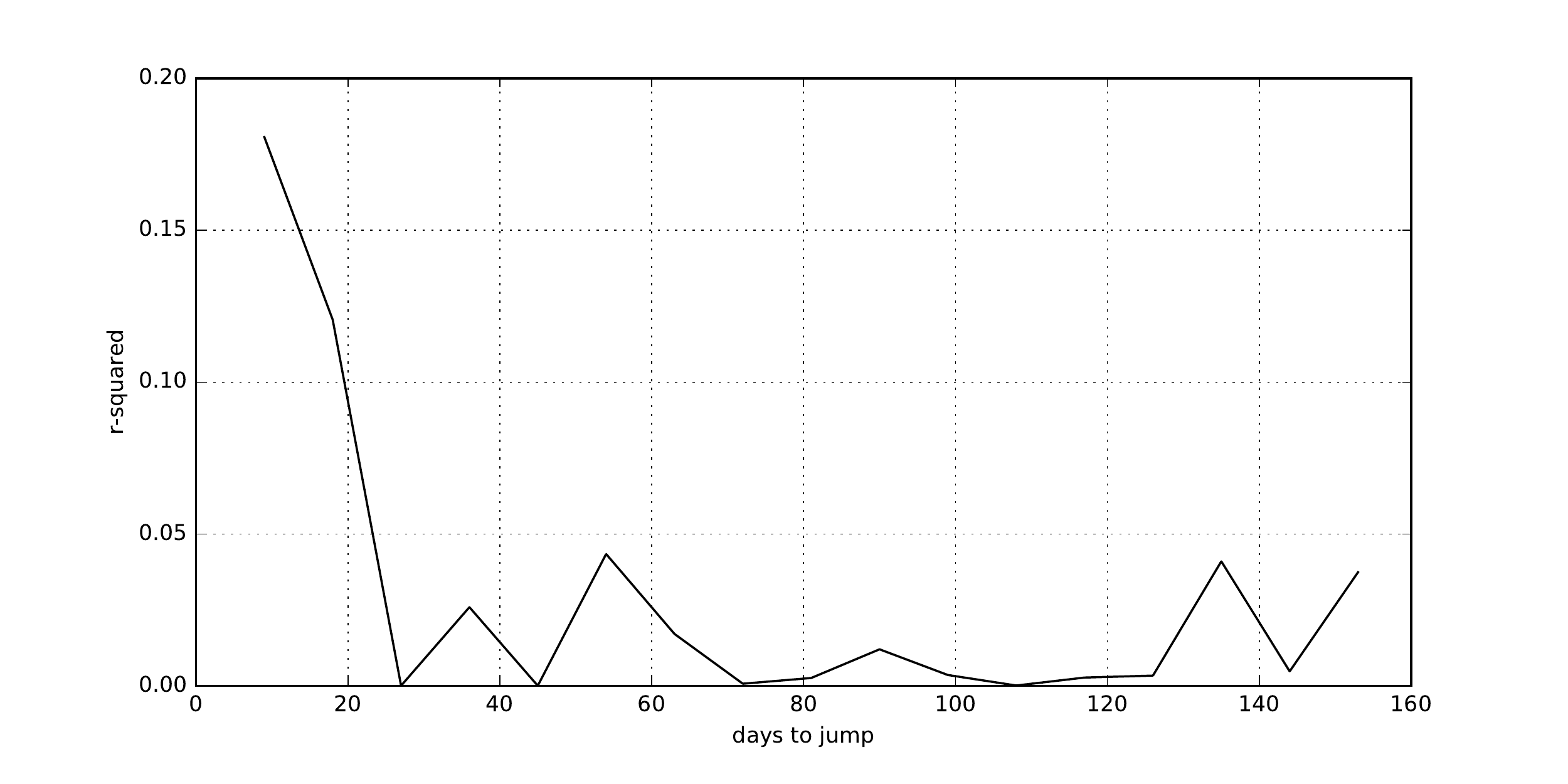}}
\caption{R-squared of SOFR realised vs forward rates for different forward periods}\label{sofr_r2}
\end{figure}

Calibration is performed for all available SOFR futures data since the commencement of trading in June 2018. The agreement between expected SOFR spikes and actual spikes is measured by calculating the R-squared between end--of--month changes in the SOFR rate $\Delta{}r(z_i)$ and the corresponding forward spike $f^Z(0,z_i)-f^Z(0,z_i-h)$. The comparison is grouped by the number of days between the calibration date corresponding to $t=0$ and $z_i$. The results in Fig. \ref{sofr_r2} show the R-squared for increasing number of days between $z_i$ and the calibration date.

The results reveal some evidence of short term anticipation of spikes close to the spike date. This is particularly true for the last trading day of the future because the trading activity in the repo market from which the day's SOFR rate is calculated occurs simultaneously to trading in the futures market. The contrast to the high R-squared for target rate jumps anticipated by Fed Funds futures comes from the fact that FOMC target rate changes are communicated well ahead of time, particularly for rate increases, while the SOFR spikes depend on liquidity conditions, which are only be anticipated in a short time frame, if at all. However, the most negative impact on the results is not lack of anticipation of spikes, rather it is the over-anticipation of spikes particularly when spikes do not occur. This indicates the presence of a spike risk premium embedded in SOFR futures prices.

It is worth noting that by focusing on the anticipation of SOFR spikes by SOFR futures, we are focusing on the \emph{incremental} information contained in SOFR futures given that FOMC target rate changes are anticipated by EFFR futures. If we were to consider SOFR futures in isolation, these also anticipate FOMC target rate changes.

\subsection{Term rate dynamics}
\begin{figure}[t]
\centerline{\includegraphics[scale=0.65]{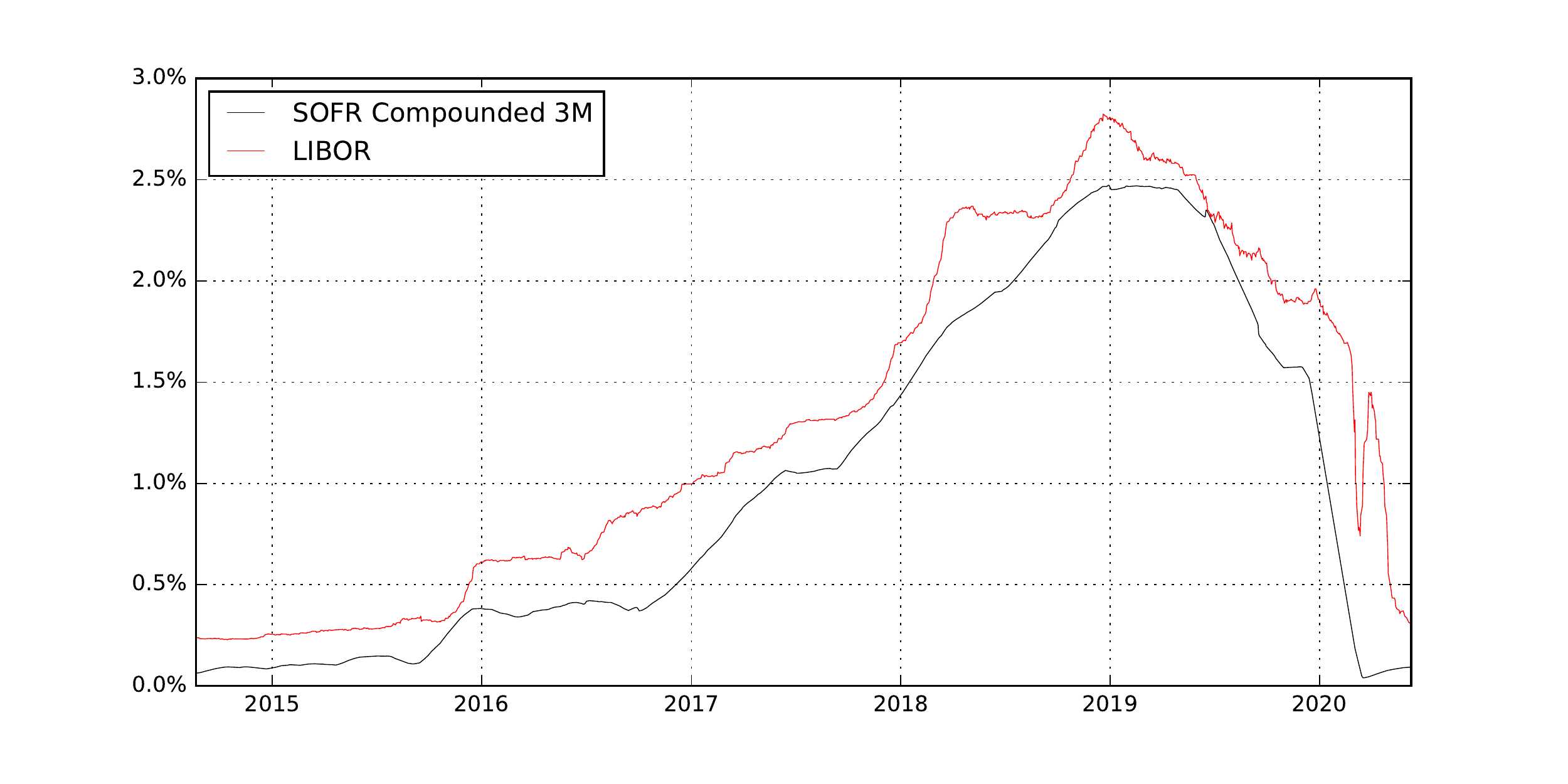}}
\caption{SOFR 3m rolling compounded rate compared to LIBOR}\label{sofr_rolling_compounded}
\end{figure}

One of the approaches considered as the replacement for the LIBOR indexation of loan terms is a rate based on retrospectively compounding of SOFR over the same term, see Fig. \ref{sofr_rolling_compounded} for a historical comparison. Both rates appear to follow the same underlying trend, this is related to the target rate term structure, which as we argue in this paper underlies all interest rates. LIBOR also exhibits considerably more volatility. This is because the SOFR compounded rate is a rolling compounding calculation of already set rates, with only one new rate rolled in the calculation on each day. LIBOR, on the other hand, is a forward looking term rate and is not subject to the volatility reduction from rolling compounding. The two rates therefore are not really comparable, which highlights one aspect of substantial problems with any proposal to replace LIBOR with a compounded SOFR.\footnote{Other problems include the disconnect due to credit risk between SOFR and the cost of funding of private--sector banks, see \citeasnoun{Berndtetal:20}.}

The calibration presented in the previous section enables a more analogous comparison of LIBOR and the SOFR forward looking spot term rate. Additionally we can examine the behaviour of SOFR 3M futures with respect to Eurodollar futures for a direct comparison of SOFR and LIBOR forward term rates. The SOFR term rates are calculated according to the compounding formula used to calculate SOFR 3M futures terminal payoff, using the daily forward rates obtained from the calibration.

\begin{figure}[t]
\centerline{\includegraphics[scale=0.65]{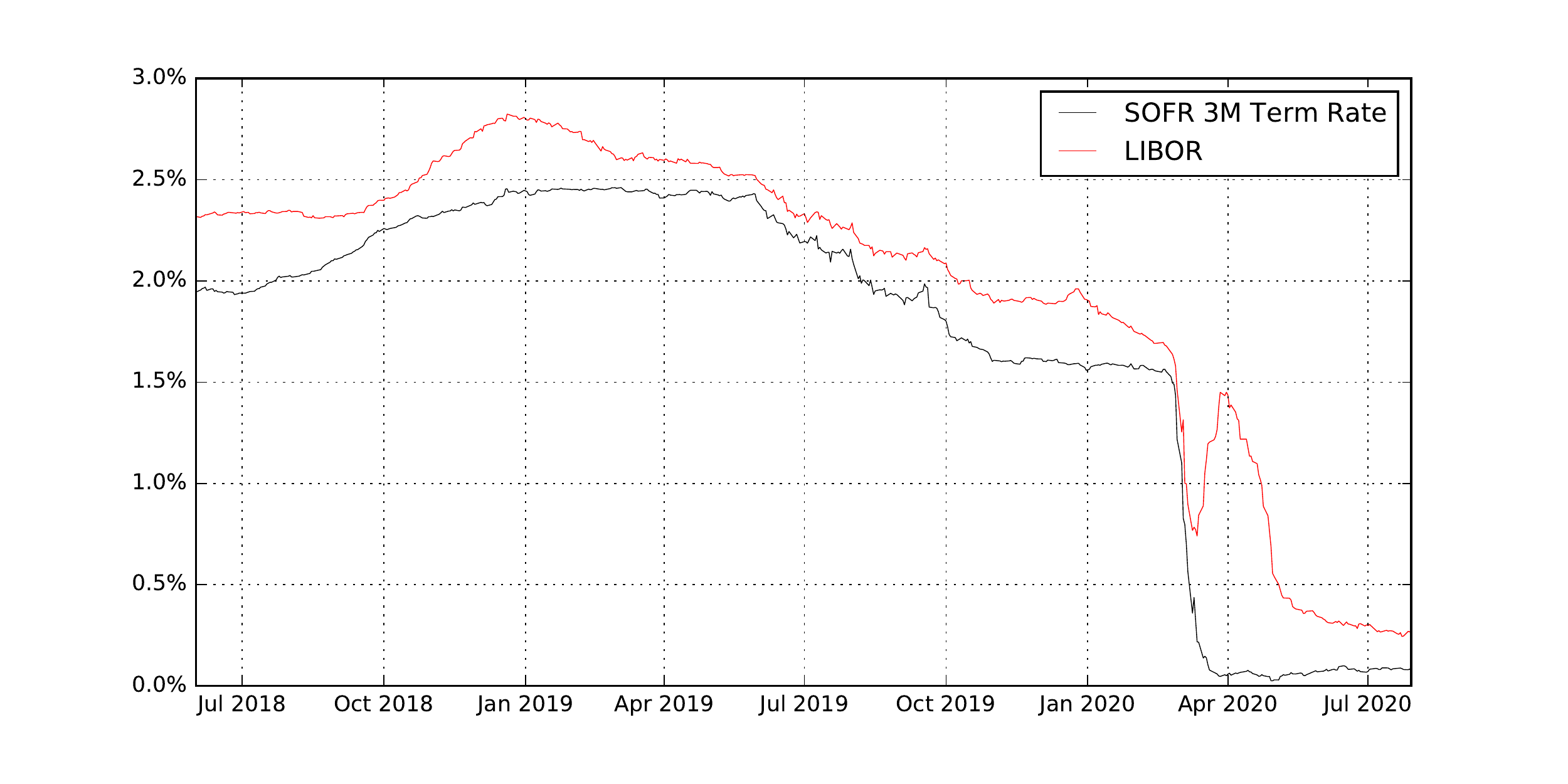}}
\caption{Spot SOFR 3m Term Rate vs LIBOR}\label{sofr_libor}
\end{figure}

The calculated spot SOFR 3M term rate is shown in Fig. \ref{sofr_libor} in comparison to spot LIBOR. The rates are well correlated, approximately 50\% of the LIBOR variance can be attributed to the SOFR 3M term rate. The impact of SOFR spikes dissipates over a 3 month compounding period, instead the term rate is mostly driven by target rate term structure. In turn, this shows that a significant proportion of LIBOR dynamics is driven by the target rate term structure exposed in our modelling framework. From this perspective, one can think of LIBOR trading at a spread to the term rates implied from the target rate term structure. One would expect this spread to be partly due to credit risk, but not entirely, since the term rate extracted from SOFR futures is not a ``true'' term rate in the sense that market participants could actually borrow at this rate\footnote{If one takes into account a borrower's risk of not being able to refinance roll--over borrowing at (a constant spread to) a benchmark rate, this gives rise to additional basis spreads as observed in the market, see \citeasnoun{Alf&Gra&Sch:18}.} --- one would therefore expect this spread also to include a ``funding liquidity risk'' component analogous to the one found in the LIBOR/OIS spread by \citeasnoun{Backwelletal:19}.

\begin{figure}[t]
\centerline{\includegraphics[scale=0.65]{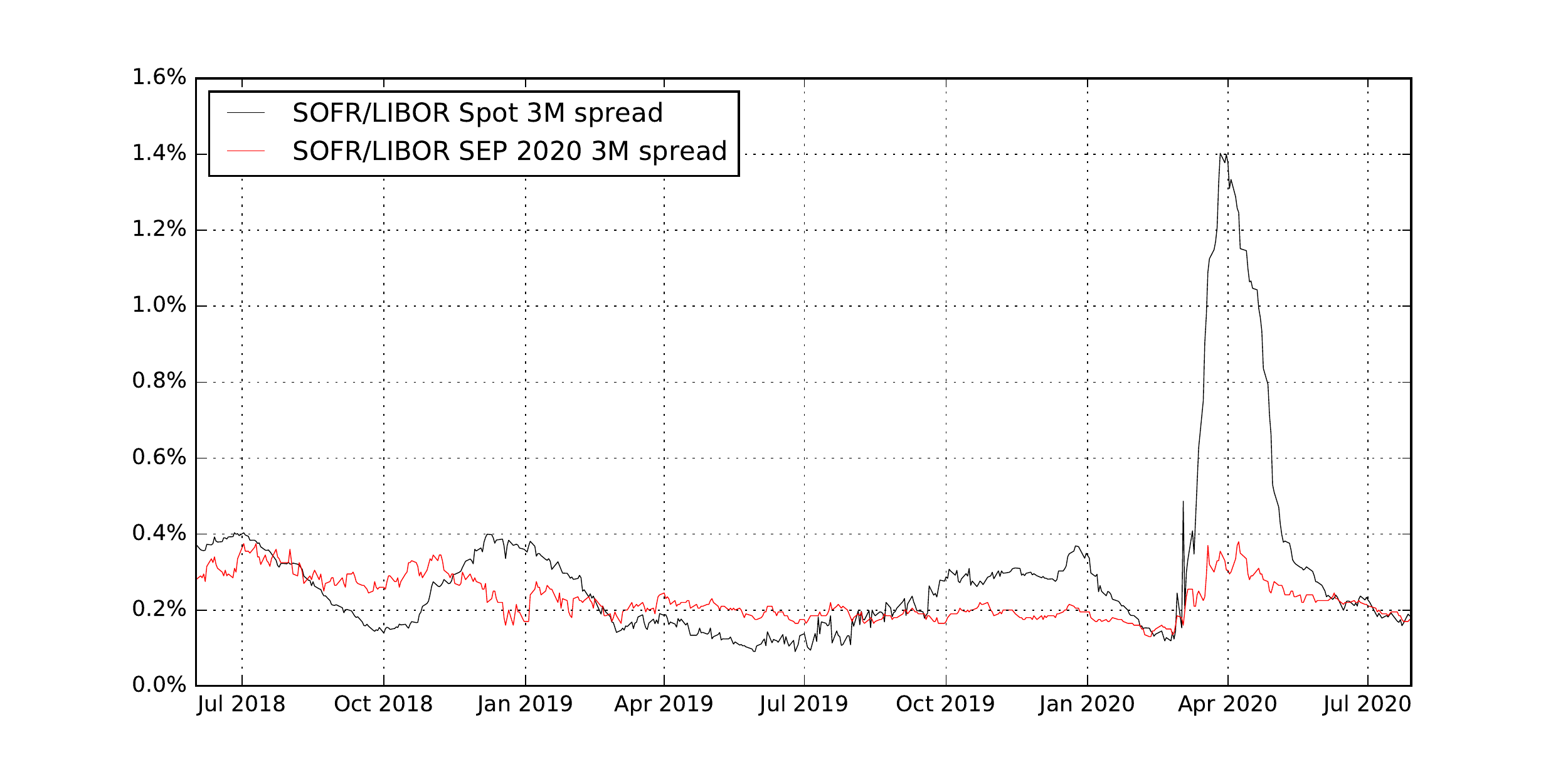}}
\caption{SOFR 3m Term Rate/LIBOR spread Spot vs Sep-2020 3M Term Rate}\label{sofr_libor_spread}
\end{figure}

It is also interesting to compare the spot and forward LIBOR to SOFR spread. As shown in Fig. \ref{sofr_libor_spread}, the spread in the forward rates appears more stable, especially during the market turmoil in February and March of 2020. This is also in contrast to the large instability exhibited by the repo rates during the financial crisis of 2008, see \citeasnoun{Andersen2020} for details. This is most likely due to Federal Reserve increasing operations in the repo market as a response to the September 2019 spike, which also appears to have eliminated end of month spikes.\footnote{See \citeasnoun{fomc} September 2019 page 5.}

\section{Conclusion}
Having observed that empirically the short rate (EFFR or SOFR) follows dynamics determined primarily by jumps at known times, but forward rates follow primarily diffusive dynamics, we have constructed a model which reconciles these two (naively contradictory) observations. Such a model is needed because, with the transition away from the LIBOR benchmark, fixed income instruments referencing SOFR are becoming increasingly important. In addition, the actions of the Federal Reserve in response to the 2008 financial crisis over the last decade have removed much of the daily volatility from the EFFR, long thought of as the best empirical proxy for the short rate. This reduction in volatility has revealed an underlying structure of short rates consisting of discontinuities directly related to FOMC policy target rate changes, which is also reflected in the empirical dynamics of SOFR. On the other hand, forward rates extracted from Fed Funds or SOFR--linked futures continue to evolve diffusively.

The model requirement, that the target rate follows a path that is constant between FOMC meeting dates, has the interesting consequence that term structure interpolation cannot be chosen arbitrarily. In particular, the popular Nelson/Siegel approach to fitting a continuous term structure to market data contradicts this requirement. Note that this is not just another manifestation of a violation of consistency in a term structure model in the sense of \citeasnoun{Bjo&Chr:99}: Piecewise constant paths of a short rate imply a no-arbitrage constraint on the shape of the initial term structure.

Calibration of our model to Fed Funds futures showed the extent (in the form of R-squared) to which these futures prices anticipate Fed Funds target rate changes. Additionally calibrating the model to SOFR futures extracts incremental, short--term information about spikes in SOFR at known times. This suggests that such spikes need to be included in the pricing and risk--management of short--term SOFR--linked instruments.

In the present paper, the driving dynamics deliberately have been kept as simple as possible. A conceptually trivial, but notationally tedious, extension would be to allow for mean reversion --- this would be necessary when considering a time horizon longer than the two years we are currently using. An extension beyond two years would also mean that the dates of FOMC meetings cannot be assumed to be precisely known (though they would still be known approximately).

Since the spikes and steps are driven by diffusive processes, the variance of spike and step magnitude depends on the length of time until the spike/step is scheduled to occur. In terms of cross--sectional calibration of the model (i.e., calibration of the model to market prices observed at a single point in time), this can be controlled by an appropriate choice of volatility functions (at present, we do not calibrate volatilities in the empirical part of the paper). Alternatively, one could modify the driving dynamics by including mean reversion and consider the model in its steady state.

For the steps in the target rate modelled in this fashion, a possible economic interpretation is that there is a fundamental ``shadow'' rate of interest evolving diffusively. Only the central bank observes this shadow rate (perhaps imperfectly), and at known dates updates the central bank target rate to match this shadow rate. As noted in Section \ref{sec:model}, our model could be extended to reflect the fact that target rates are only updated in discrete increments.

\begin{appendices}
\section{}
For a Brownian motion $W(t)$:
\begin{gather}
\int\limits_0^t\mathbbm{1}(s<x)dW(s)=W(t\wedge{}x)
\end{gather}
Therefore we have the following solutions to various stochastic integrals appearing in this paper:
\begin{gather}\label{appendix_A1}
	\int\limits_0^t\xi_i\mathbbm{1}(s<x_i)\mathbbm{1}(T\geq{}x_i)dZ_i(s)=\xi_i\mathbbm{1}(T\geq{}x_i)Z_i(t\wedge{}x_i)
\end{gather}
\begin{gather}\label{appendix_A2}
	\int\limits_0^t\xi_i\mathbbm{1}(s<x_i)\mathbbm{1}(t\geq{}x_i)dZ_i(s)=\xi_i\mathbbm{1}(t\geq{}x_i)Z_i(x_i)
\end{gather}
\begin{gather}\label{appendix_A3}
	\int\limits_0^t\sum\limits_{i=1}^n\xi_i\lambda_{i,j}\mathbbm{1}(s<x_i)\mathbbm{1}(T\geq{}x_i)dW^P_j(s)=\sum\limits_{i=1}^n\xi_i\lambda_{i,j}\mathbbm{1}(T\geq{}x_i)W^P_j(t\wedge{}x_i)
\end{gather}
\begin{gather}\label{appendix_A6}
	\int\limits_0^t\sum\limits_{i=1}^n\xi_i\lambda_{i,j}\mathbbm{1}(s<x_i)\mathbbm{1}(t\geq{}x_i)dW^P_j(s)=\sum\limits_{i=1}^n\xi_i\lambda_{i,j}\mathbbm{1}(t\geq{}x_i)W^P_j(x_i)
\end{gather}
\begin{gather}\label{appendix_A9}
	\int\limits_0^t\sigma^Z_i\mathbbm{1}(s<z_i)\mathbbm{1}(T\in{}H_i)dW^Z_i(s)=\sigma^Z_i\mathbbm{1}(T\in{}H_i)W^Z_i(t\wedge{}z_i)
\end{gather}
\begin{gather}\label{appendix_A12}
	\int\limits_0^t\sigma^Z_i\mathbbm{1}(s<z_i)\mathbbm{1}(t\in{}H_i)dW^Z_i(s)=\sigma^Z_i\mathbbm{1}(t\in{}H_i)W^Z_i(z_i)
\end{gather}
Solving the drift term $\int\limits_0^t\sigma^P_j(u,T)\int\limits_u^T\sigma^P_j(u,s)dsdu$, we have:
\begin{align}
\begin{split}
\int\limits_u^T\sigma^P_j(u,s)ds&=\int\limits_u^T\sum\limits_{i=1}^n\xi_i\lambda_{i,j}\mathbbm{1}(u<x_i)\mathbbm{1}(s\geq{}x_i)ds\\
&=\sum\limits_{i=1}^n\xi_i\lambda_{i,j}\mathbbm{1}(u<x_i)\int\limits_u^T\mathbbm{1}(s\geq{}x_i)ds\\
\end{split}
\end{align}
where:
\begin{gather}\label{eq:Cases}
   \int\limits_u^T\mathbbm{1}(s\geq{}x_i)ds=
	\begin{cases}
		0,&T<x_i\\
		T-x_i,&u<x_i,T\geq{}x_i\\
		T-u,&u\geq{}x_i
   \end{cases}
\end{gather}
therefore:
\begin{gather}\label{appendix_A4}
   \int\limits_u^T\sigma^P_j(u,s)ds=\sum\limits_{i=1}^n\xi_i\lambda_{i,j}\mathbbm{1}(u<x_i)\mathbbm{1}(T\geq{}x_i)(T-x_i)
\end{gather}
therefore:
\begin{align}\label{appendix_A5}
\begin{split}
&\int\limits_0^t\sigma^P_j(u,T)\int\limits_u^T\sigma^P_j(u,s)dsdu\\
&=\int\limits_0^t\sum\limits_{q=1}^n\xi_q\lambda_{q,j}\mathbbm{1}(u<x_q)\mathbbm{1}(T\geq{}x_q)\sum\limits_{i=1}^n\xi_i\lambda_{i,j}\mathbbm{1}(u<x_i)\mathbbm{1}(T\geq{}x_i)(T-x_i)du\\
&=\sum\limits_{q=1}^n\xi_q\lambda_{q,j}\mathbbm{1}(T\geq{}x_q)\sum\limits_{i=1}^n\xi_i\lambda_{i,j}\mathbbm{1}(T\geq{}x_i)(T-x_i)\int\limits_0^t\mathbbm{1}(u<x_q)\mathbbm{1}(u<x_i)du\\
&=\sum\limits_{q=1}^n\sum\limits_{i=1}^n\xi_q\xi_i\lambda_{q,j}\lambda_{i,j}\mathbbm{1}(T\geq{}x_{q\vee{}i})(T-x_i)[t\wedge{}x_q\wedge{}x_i]
\end{split}
\end{align}
Similarly:
\begin{gather}\label{appendix_A8}
\int\limits_0^t\sigma^P_j(u,t)\int\limits_u^t\sigma^P_j(u,s)dsdu=\sum\limits_{q=1}^n\sum\limits_{i=1}^n\xi_q\xi_i\lambda_{q,j}\lambda_{i,j}\mathbbm{1}(t\geq{}x_{q\vee{}i})(t-x_i)[x_q\wedge{}x_i]
\end{gather}
Solving $\int\limits_0^t\sigma^Z_i(u,T)\int\limits_u^T\sigma^Z_i(u,s)dsdu$:
\begin{align}\label{appendix_A10}
\begin{split}
\int\limits_u^T\sigma^Z_i(u,s)ds&=\int\limits_u^T\sigma^Z_i\mathbbm{1}(u<z_i)\mathbbm{1}(s\in{}H_i)ds\\
&=\sigma^Z_i\mathbbm{1}(u<z_i)\int\limits_u^T\mathbbm{1}(s\in{}H_i)ds\\
&=\sigma^Z_i\mathbbm{1}(u<z_i)\mathbbm{1}(T\geq{}z_i)[h_i\wedge{}(T-z_i)]
\end{split}
\end{align}
therefore
\begin{align}\label{appendix_A11}
\begin{split}
&\int\limits_0^t\sigma^Z_i(u,T)\int\limits_u^T\sigma^Z_i(u,s)dsdu\\
&=\int\limits_0^t\sigma^Z_i\mathbbm{1}(u<z_i)\mathbbm{1}_{H_i}(T)\sigma^Z_i\mathbbm{1}(u<z_i)\mathbbm{1}(T\geq{}z_i)[h_i\wedge{}(T-z_i)]du\\
&=\big(\sigma^Z_i\big)^2\mathbbm{1}(T\in{}H_i)[h_i\wedge{}(T-z_i)]\int\limits_0^t\mathbbm{1}(u<z_i)du\\
&=\big(\sigma^Z_i\big)^2\mathbbm{1}(T\in{}H_i)(T-z_i)[t\wedge{}z_i]\\
\end{split}
\end{align}
Similarly:
\begin{gather}\label{appendix_A14}
\int\limits_0^t\sigma^Z_i(u,t)\int\limits_u^t\sigma^Z_i(u,s)dsdu=\big(\sigma^Z_i\big)^2\mathbbm{1}(t\in{}H_i)(t-z_i)z_i
\end{gather}

\end{appendices}

\bibliography{SOFR_BIB}

\end{document}